\documentclass[pdftex,a4paper,titlepage]{scrartcl}
\pdfoutput=1
\usepackage[utf8]{inputenc}
\usepackage[T1]{fontenc}
\usepackage{lmodern}
\usepackage{microtype}
\usepackage[english,ngerman]{babel}
\usepackage[babel]{csquotes}
\usepackage[pdftex]{graphicx}
\usepackage{lipsum}
\usepackage{placeins}
\usepackage{verbatim}
\usepackage{color}
\usepackage{amssymb}
\usepackage{amsmath}

\addtokomafont{disposition}{\boldmath}

\usepackage[numbers]{natbib}

\usepackage[pdftex]{hyperref}
\hypersetup{colorlinks=true,
linkcolor=black,
citecolor=black,
filecolor=black,
urlcolor=black,
pdftitle={Bachelorarbeit Quantenzufallsgenerator -- Bastian Hacker},
pdfauthor={Bastian Hacker},
pdfsubject={Entwicklung eines schnellen optischen Quanten-Zufallsgenerators},
pdfcreator={pdflatex},
pdfkeywords={Quantum Random Number Generator, Quanten-Zufallsgenerator}}

\newcommand{\degr}{\ensuremath{^\circ}}
\newcommand{\dif}{\ensuremath{\mathrm{d}}}

\newcommand{\unit}[1]{\ensuremath{\,\mathrm{#1}}}
\newcommand{\ee}[1]{\ensuremath{\cdot 10^{#1}}}

\addtokomafont{caption}{\small}
\addtokomafont{captionlabel}{\small}

\title{Entwicklung eines schnellen optischen Quanten-Zufallsgenerators}
\author{Bastian Hacker}
\date{23. September 2011}

\begin{document}
\begin{titlepage}
\mbox{}\vspace{2cm}
\begin{center}
\Large{
{\titlefont\Huge Entwicklung eines schnellen optischen Quanten-Zufallsgenerators}
\par\vskip 3em
\textbf{Bachelorarbeit}\\
im Fach Physik
\par\vskip 2em
Bastian Hacker
\par\vskip 1em
22.\,9.\,2011
\par\vskip 6em
Betreuer: Prof. Dr. G. Leuchs
\par\vskip 1em
Arbeitsgruppe Quanteninformationsverarbeitung (QIV)
\par\vskip 1em
Max-Planck-Institut für die Physik des Lichts,\\
Friedrich-Alexander-Universität Erlangen-Nürnberg}
\end{center}
\vfill
\noindent
\begin{minipage}[b]{\textwidth}
\includegraphics[height=46pt]{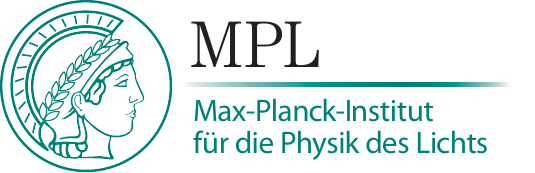}
\hfill
\raisebox{6pt}{\includegraphics[height=30pt]{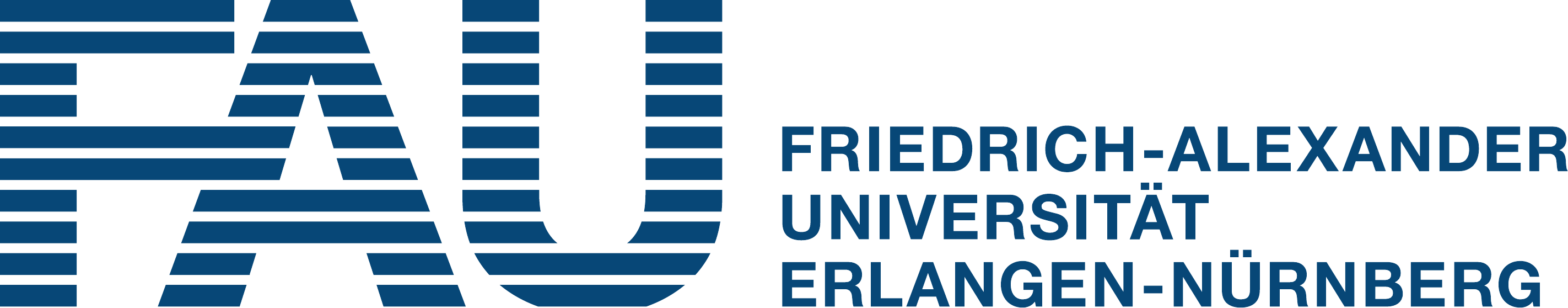}}
\end{minipage}
\end{titlepage}

\sloppy
\thispagestyle{empty}
\let\raggedsection\centering
\section*{\abstractname}
Diese Arbeit behandelt Aufbau, Charakterisierung und Datenverarbeitung eines echten Quanten-Zufallszahlengenerators.
Als Zufallsquelle dient ein reiner Quanten-Vakuumzustand, der mit einem Laserstrahl verstärkt wird.
Leistungsfähigkeit und Verhalten des Systems so wie mögliche Störeinflüsse werden untersucht.
Zur optimalen Nutzung der Rohdaten werden diese Fourier-transformiert und als Frequenzamplituden weiterverarbeitet.
In diesen Daten wird die extrahierbare Entropie bestimmt, um eine Elimination nicht-zufälliger Signalbeiträge durch Hashing zu ermöglichen.
Das System kann echte und einzigartige Zufallszahlen mit einer Rate von $25\unit{Gbit/s}$ erzeugen und übertrifft damit bisherige Realisierungen deutlich.

\selectlanguage{english}
\section*{\abstractname}
\let\raggedsection\raggedright
This work reports on setup, characterisation and data processing of a true quantum random number generator.
As a randomness source a pure quantum vacuum state of light is used, which is amplified by a laser beam.
Performance and behaviour of the system as well as parasitic errors are investigated.
For an optimized exploitation of the data they are Fourier-transformed and processed further as frequency amplitudes.
The extractable entropy in that data is calculated to allow for elimination of non-random signal contributions by hashing.
The system is able to produce true and unique random numbers at a rate of $25\unit{Gbit/s}$ and thus outperforms previous implementations considerably.

\selectlanguage{ngerman}

\newpage
\tableofcontents

\newpage
\section{Einleitung}

\begin{quote}
"`Das, wobei unsere Berechnungen versagen, nennen wir Zufall."'\\
\mbox{}\hfill \emph{Albert Einstein}
\end{quote}

Als Zufall bezeichnen wir Ereignisse, die durch keine Systematik vollständig bestimmt sind.
Dies erscheint oftmals so aufgrund von fehlendem Wissen über prinzipiell fest bestimmte Details.
So beispielsweise bei der erzielten Augenzahl eines Würfels, dessen Bewegung mit bekannten Gesetzen genau vorhergesagt werden könnte.
Im Gegensatz dazu soll ein derartiger Determinismus bei echtem Zufall ausgeschlossen sein.

\subsection{Zufallszahlen}
Um den Zufall wissenschaftlich zu fassen, werden diskrete oder kontinuierliche Zufallszahlen eingeführt, also Zahlen die einen zufälligen Wert aus einer gegebenen Menge annehmen können.
In dieser Form lässt sich der Zufall analysieren und quantifizieren, speichern und anwenden.

Solche Zufallszahlen sind zu einem wichtigen Werkzeug in unserer heutigen Informationsgesellschaft geworden.
Eigenschaften wie Unvorhersagbarkeit oder bestimmtes statistisches Verhalten bieten einen oftmals unverzichtbaren Nutzen.
Bei Glücksspielen ist der Einsatz offensichtlich.
Viele Computeralgorithmen sind aber ebenfalls auf Zufallszahlen angewiesen, obwohl ein klar definiertes nicht-zufälliges Ergebnis errechnet werden soll.
Das kann in Simulationen sein um Spezialfälle zu vermeiden und typische, "`zufällige"' Situationen zu erzeugen, oder in Fällen wo zufällige Parameter im Durchschnitt schneller zum Ziel führen als klug ausgewählte~\citep{motwani1995randomized}.
Hierfür müssen die Zahlen schnell genug erzeugt werden um das Programm nicht auszubremsen, die völlige Unvorhersagbarkeit ist dagegen nicht entscheidend.
Anders bei kryptographischen Anwendungen.
Zur Verschlüsselung geheimer Nachrichten darf ein potentieller Abhörer nichts über den verwendeten Schlüssel wissen.
Dieser muss also ein sehr hohes Maß an Zufälligkeit besitzen.

\subsection{Zufallsgeneratoren}
Die Erzeugung von Zufallszahlen ist keine triviale Aufgabe, denn Zufall entsteht nach den klassischen Regeln der Informationstheorie nicht einfach.
Innerhalb von deterministischen Systemen wie einem klassischen Computer ist es gar unmöglich eine zufällige Ausgabe zu erzeugen.
Wohl aber gibt es eine ganze Reihe von Pseudozufallsgeneratoren, die deterministische Zahlenfolgen mit vielen zufallsähnlichen Eigenschaften erzeugen können.
Solche Generatoren sind in Form von Computerprogrammen sehr einfach zu handhaben und effizient \citep{matsumoto1997mt}.
Die Ergebnisse sind aber niemals echt zufällig und mit gegebenen Startwerten sogar direkt reproduzierbar.

Besser sind deshalb physikalische Zufallszahlengeneratoren.
Diese messen ein physikalisches System mit zufälligen Eigenschaften, beispielsweise thermisches Rauschen.
Wegen der großen Zahl an Freiheitsgraden sind solche Systeme praktisch nicht vorhersagbar und bieten gute Quellen für Zufallszahlen.
Eine kleine Unsicherheit verbleibt aber auch hier.
Physikalische Systeme sind modellierbar, und mit zukünftigen Rechenleistungen könnte es prinzipiell möglich werden, beliebig komplexen Systemen ihre Zufälligkeit zu nehmen.

\subsection{Quanten-Zufallsgeneratoren}
Der ultimative Zufallsgenerator sollte deshalb auf einem intrinsisch zufälligen System basieren, das nicht nur durch unvollständiges Wissen zufällig erscheint, sondern nicht-kausalen Naturgesetzen gehorcht.
Die Existenz solcher Systeme ist nicht selbstverständlich.
Mit der Quantenmechanik existiert jedoch eine experimentell exzellent gestützte Theorie, die intrinsischen Zufall als zentrales Element enthält.
Gemäß der Kopenhagener Deutung der Quantenmechanik wird der Wert einer Messgröße, die sich nicht im entsprechenden Eigenzustand befindet, beim Messvorgang entsprechend einer Wahrscheinlichkeitsverteilung echt zufällig ausgewählt.

Als geeignete Quantensysteme haben sich hierfür Photonen, also Lichtteilchen, herauskristallisiert, weil diese sich mit hohen Raten erzeugen und messen lassen.
Auf diesem Gebiet gab es besonders in der letzten Dekade gewaltige Fortschritte, z.B. in der Erzeugung und Detektion quantenmechanischer Zustände.
Mögliche Ansätze sind die Messung der Ankunftszeiten kohärenter Einzelphotonen \citep{Jennewein2000, dynes2008high} oder spontane Emission \citep{stipcevic2007quantum}.

Ein besonders vielversprechender Ansatz, der in dieser Arbeit verfolgt wurde, ist die Messung vom Schrotrauschen eines reinen Vakuumzustandes mit Hilfe eines kontinuierlichen Laserstrahls \citep{Gabriel2010}.
Hierbei hat man eine unbeeinflussbare Zufallsquelle und einen übersichtlichen Messaufbau.
Im folgenden wird gezeigt werden, dass damit auch sehr hohe Datenraten erreicht werden können.

Ein besonderes Augenmerk wird dabei auf die Qualität der Zufallszahlen gelegt werden.
Aufgrund klassischer Fluktuationen kann man mit keiner Apparatur nur Quantenrauschen -- d.\,h. echten Zufall -- messen, da immer klassische und damit deterministische Signale das Messsignal beeinflussen werden.
Die meisten Ansätze verwenden deshalb problemspezifische Korrekturen möglicher Verzerrungen \citep{Furst2010,Shen2010}.
Dieses Projekt geht aber einen Schritt weiter.
Hier wird die Menge an extrahierbarem Zufall in den aufbereiteten Zufallszahlen berechnet, so dass ein generisches mathematisches Verfahren deren Zufälligkeit auf einen kleineren Satz von finalen Zufallszahlen konzentrieren kann.
Das wurde erstmals im Vorgängerprojekt demonstriert \citep{Gabriel2010}.
Hieraus resultieren echte Zufallszahlen, die allen kryptographischen Anforderungen genügen.

\newpage
\FloatBarrier
\section{Theorie}
\begin{figure}
	\centering
		\includegraphics[width=\textwidth]{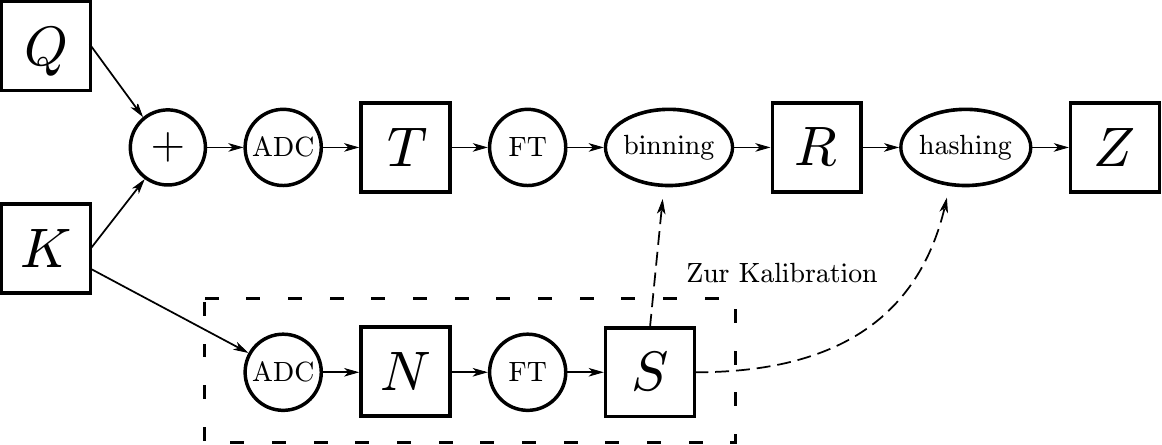}
	\caption{Informationsfluss: Zum Quantenrauschsignal $Q$ wird beim Messen klassisches Rauschen $K$ addiert.
	Das Gesamtsignal wird im analog-digital Wandler (ADC) des Oszilloskops zum Signal $T$ digitalisiert, danach fouriertransformiert (FT) und zu Roh-Zufallszahlen $R$ gebinnt.
	Später werden die Zufallszahlen $Z$ durch Hashing extrahiert.
	Zur Kalibration wird das Dunkelrauschen $K$ direkt gemessen und zu $N$ digitalisiert, deren Fouriertransformierte $S$ dazu dient, Binning und Hashingverhältnis richtig einzustellen.}
	\label{fig:infoflow}
\end{figure}

Das Experiment in dieser Arbeit nutzt einen Laserstrahl, über den das Quanten\-rauschen des Vakuums messbar gemacht wird.
Das Signal in Form von Spannungswerten am Detektor wird von einem Oszilloskop digitalisiert und gespeichert (siehe Abb.~\ref{fig:infoflow}).

Soweit enthält das Signal zusätzliches Elektronikrauschen und zeitliche Korrelationen durch den Frequenzgang von Detektor und Oszilloskop.
Hier setzt die Kernidee dieser Arbeit an:
Die zeitlichen Messwerte werden mit einer Fouriertransformation stückweise in den Frequenzraum umgerechnet (Kap.~\ref{kap:ft}).
Dort verschwinden die störenden Korrelationen zwischen den Messwerten weitgehend (Kap.~\ref{sec:theospec}) und äußern sich lediglich durch schwächere Amplituden bei höheren Frequenzen.
Elektronikrauschen und Störeinflüsse beeinflussen die verschiedenen Komponenten unterschiedlich stark und können angepasst behandelt werden.

Es ist wichtig zu wissen, wie unabhängig verschiedene Frequenzkomponenten tatsächlich sind, wie gut die Messsignale und wie groß die Auswirkungen von Störeinflüssen.
Dazu werden in Kap.~\ref{sec:messung} zahlreiche Eigenschaften des Systems vermessen.

Zur Erzeugung der letztendlichen Zufallszahlen werden die berechneten Frequenzamplituden gebinnt, also in gröbere Einheiten digitalisiert und zwar so, dass die resultierenden Werte einer zufälligen Gleichverteilung möglichst nahe kommen (Kap.~\ref{kap:binning}).
Im letzten Schritt werden die Werte mit einer kryptographischen Hashfunktion auf eine kleinere Menge finaler Zufallszahlen abgebildet.
Das Hashing selbst war nicht Teil dieser experimentellen Arbeit.
Allerdings wird gezeigt, welche Hashingrate nötig sein wird, um echte Zufallszahlen zu erhalten, selbst unter der Annahme dass der Verlauf des Elektronikrauschens vollständig bekannt werden könnte (Kap.~\ref{sec:extraktion}).

\subsection{Quantenmechanische Beschreibung der Lichtzustände}
\label{kap:quantenstatistik}
In diesem Experiment nutzen wir die Quantennatur von Licht und die daraus resultierende Verteilung des Detektorsignals.
Licht wird in der Quantenmechanik in Zustandsvektoren beschrieben, beispielsweise $|n\rangle$ für einen Zustand mit definierter Photonenzahl $n$ in einer bestimmten Frequenzmode.
Licht mit bestmöglich definierter Feldamplitude und Phase kann als kohärenter Zustand $|\alpha\rangle$ geschrieben werden, wobei $|\alpha|^2=\bar n$ die mittlere Photonenzahl und $\mathrm{arg}(\alpha)$ die mittlere Phase der Welle ist.
Die messbare Photonenzahl und Intensität eines solchen Zustandes genügt dann der Wahrscheinlichkeitsverteilung \cite{glauber1963coherent}
\begin{equation}
P_\alpha(n) = |\langle n|\alpha\rangle|^2
= \exp\left(-\bar n\right)\frac{{\bar n}^{n}}{n!}
\ ,
\end{equation}
also einer Poisson-Verteilung, die der von unabhängig ankommenden Photonen entspricht.

Die Poisson-Verteilung des Quantenrauschens besitzt eine Varianz $\sigma^2(n)=\bar n$.
Klassisches Rauschen dagegen skaliert linear mit der Intensität und besitzt daher die Varianz $\sigma^2(n)={\bar n}^2$.
Diese Eigenschaft wird in dieser Arbeit zum experimentellen Nachweis des Quantenrauschens genutzt.

\FloatBarrier
\subsection{Homodyn-Detektion}
\label{sec:homodyn}
Ein kohärenter Laserstrahl wäre so bereits als Zufallsquelle nutzbar.
Allerdings ließe sich die Intensität am Detektor durch äußere Einflüsse wie z.\,B. hochfrequente Schwankungen der Netzspannung direkt manipulieren und noch schlimmer,
es wäre eine Verschränkung des Strahls denkbar, die Informationen über seinen Quantenzustand nach außen trägt.

\begin{figure}[t!]
	\centering
		\includegraphics[width=\textwidth]{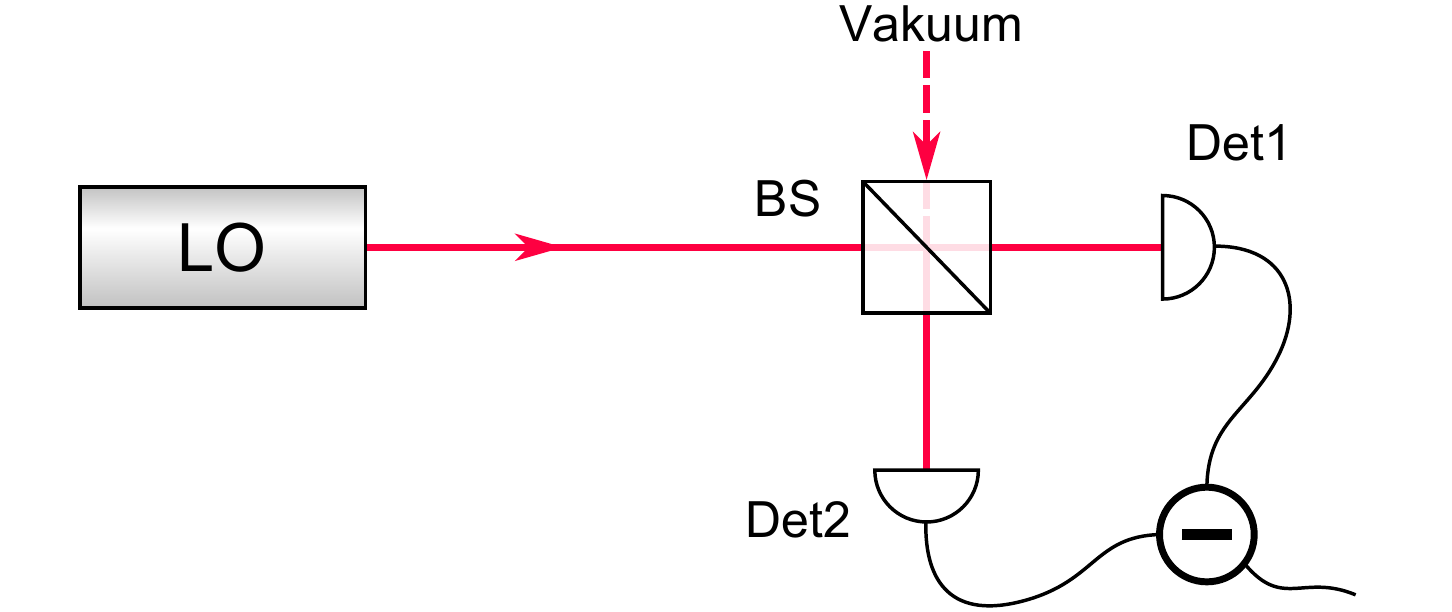}
	\caption{Schema der Homodyn-Detektion. Ein Referenzstrahl (LO) wird am Strahlteiler (BS) mit dem Vakuumzustand interferiert.
Aus der Differenz der entstehenden Teilstrahlen kann hiernach eine verstärkte Version des Vakuumzustandes gemessen werden.}
	\label{homodyn_skizze}
\end{figure}

Deshalb wurde hier der sogenannte Homodyn-Aufbau gewählt (Abb. \ref{homodyn_skizze}).
Dabei wird eine Referenzquelle (lokaler Oszillator\footnote{Die Bezeichnung lokaler Oszillator ist aus der Funktechnik übernommen, wo das gleiche Messprinzip mit einem empfängerseitigen Oszillator realisiert wird.}, LO) mit einem zu messenden Signal in einem 50:50 Strahlteiler (BS\footnote{Aus dem Englischen \emph{beam splitter}}) zur Interferenz gebracht, beide Teilstrahlen einzeln detektiert und das Differenzsignal aufgezeichnet.
Klassisch wäre zu erwarten, dass das Differenzsignal im Idealfall verschwindet.

Das ist allerdings nicht der Fall:
Der Strahlteiler besitzt nämlich einen zweiten Eingang.
Auch wenn dieser geblockt ist und kein Licht hinein gelangt, so liegt dort immer noch der Vakuumzustand\footnote{Das Experiment befindet sich nicht im Vakuum sondern in Luft. Der Zustand ist aber danach benannt, weil er selbst im Vakuum noch vorhanden wäre.} an und interferiert mit dem LO zu zwei verschiedenen Teilstrahlen \cite{gerry2005introductory}.
Das resultierende Differenzsignal ist damit eine durch den LO verstärkte Version des Vakuumzustandes, genauer gesagt, je nach Phasenlage des LO, einer der Quadraturen $\hat X$ oder $\hat P$ (entspricht Sinus- und Kosinusanteil).
Die Phasenlage des Vakuumzustandes ist aber beliebig, weshalb man annehmen kann die $\hat X$-Quadratur zu messen.
Diese $\hat X$-Quadratur besitzt nun wieder die Statistik eines kohärenten Zustandes wie in Kap.~\ref{kap:quantenstatistik} die sich für hohe LO Intensitäten einer Normalverteilung annähert.
Fluktuationen im LO selbst beeinflussen das Mess\-signal nun in erster Näherung nicht mehr \cite{bachor2004guide}.

\subsection{Fouriertransformation}
\label{kap:ft}
Im Idealfall sollte aus der Messung des Vakuumzustandes ein zufälliges, zeitlich unkorreliertes Signal erzeugt werden.
Im realen System steht aber nur das Ausgangssignal des Oszilloskops zur Verfügung, in dem das Messsignal frequenzabhängig verstärkt wird und dem das Elektronikrauschen des Detektorsystems überlagert ist.
Zur Trennung dieser Effekte betrachten wir das Messsignal stückweise im Frequenzraum, das heißt für jeweils eine endliche Zahl an Datenpunkten wird die diskrete Fouriertransformierte (DFT) berechnet
\begin{equation}
\textnormal{DFT}:\quad
\underline a=(a_0, \ldots, a_{N-1}) \in \mathbb{C}^N
\quad\rightarrow\quad
\tilde{\underline{a}}=(\tilde{a}_0, \ldots, \tilde{a}_{N-1}) \in \mathbb{C}^N
\nonumber
\end{equation}
\begin{equation}
\tilde{a}_k = \frac{1}{\sqrt{N}}\cdot
\sum_{j=0}^{N-1} a_j \cdot e^{-2\pi\mathrm{i}\cdot\frac{jk}{N}}
\label{eq:fourier}
\end{equation}
\begin{equation}
\textnormal{mit}\quad
a_j = a\bigl(\,t = t_0 + j\cdot \Delta t\,\bigr)
\quad,\quad
\tilde{a}_k = \tilde{a}\bigl(\,f = \frac{k}{N\cdot\Delta t}\,\bigr)
\nonumber\ ,
\end{equation}
die angibt, aus welchen Frequenzkomponenten sich das Signal zusammensetzen lässt.
Rauschen durch Driften des Signals, Schallwellen, Vibrationen und ähnliches bleibt in den entsprechenden niederen Frequenzkomponenten konzentriert.
Die begrenzte Bandbreite des Detektors dagegen kommt erst bei hohen Frequenzkomponenten zum Tragen.
Für jede Komponente kann nun ihre mittlere Signalamplitude und das mittlere Dunkelrauschen ermittelt werden, was zur Berechnung der extrahierbaren Signalinformation nötig ist.

Die DFT arbeitet mit komplexen Zahlen.
Ist das Eingangssignal aber reell, $a \in \mathbb{R}^N$, so wird  die Ausgabe hermitesch $\tilde{a}_{N-k} = \overline{\tilde{a}_k}$ und die zweite Hälfte der Transformierten ist redundant.
Die Komponenten $(\tilde{a}_1, \ldots, \tilde{a}_{N/2-1})$ werden durch die Symmetrie nicht eingeschränkt, so dass wir deren Real- und Imaginärteil jeweils als unabhängige Größen verwenden können.

Zur Charakterisierung der Fouriertransformierten wird die spektrale Leistungsdichte hergenommen, also der quadratisch gemittelte Wert aus Real- und Imaginärteil
\begin{equation}
S(f) = \langle |\tilde a(f)|^2\rangle_t
\ .
\end{equation}
Das so definierte Spektrum besitzt für endliche Frequenzauflösungen endliche Werte.
Weil spektrale Leistungsdichten oft über viele Größenordnungen variieren, wird gerne die Dezibelskala
\begin{equation}
S\unit{[dB]} = 10\cdot\log_{10}\left( S/S_0 \right)
\end{equation}
verwendet.
$S_0$ kann hier je nach Signaleinheit beispielsweise $1\unit{V^2}$ sein.

Allerdings kann die DFT mit endlichen Zeitabschnitten das Spektrum nicht beliebig genau bestimmen.
Dies liegt an der mathematischen Zeit-Frequenz-Unschärfe, hier in Form des "`Leck-Effektes"'.

\FloatBarrier
\subsubsection{Leck-Effekt}
\label{kap:leck-effekt}
\begin{figure}[t!]
	\centering
		\includegraphics[width=1\textwidth]{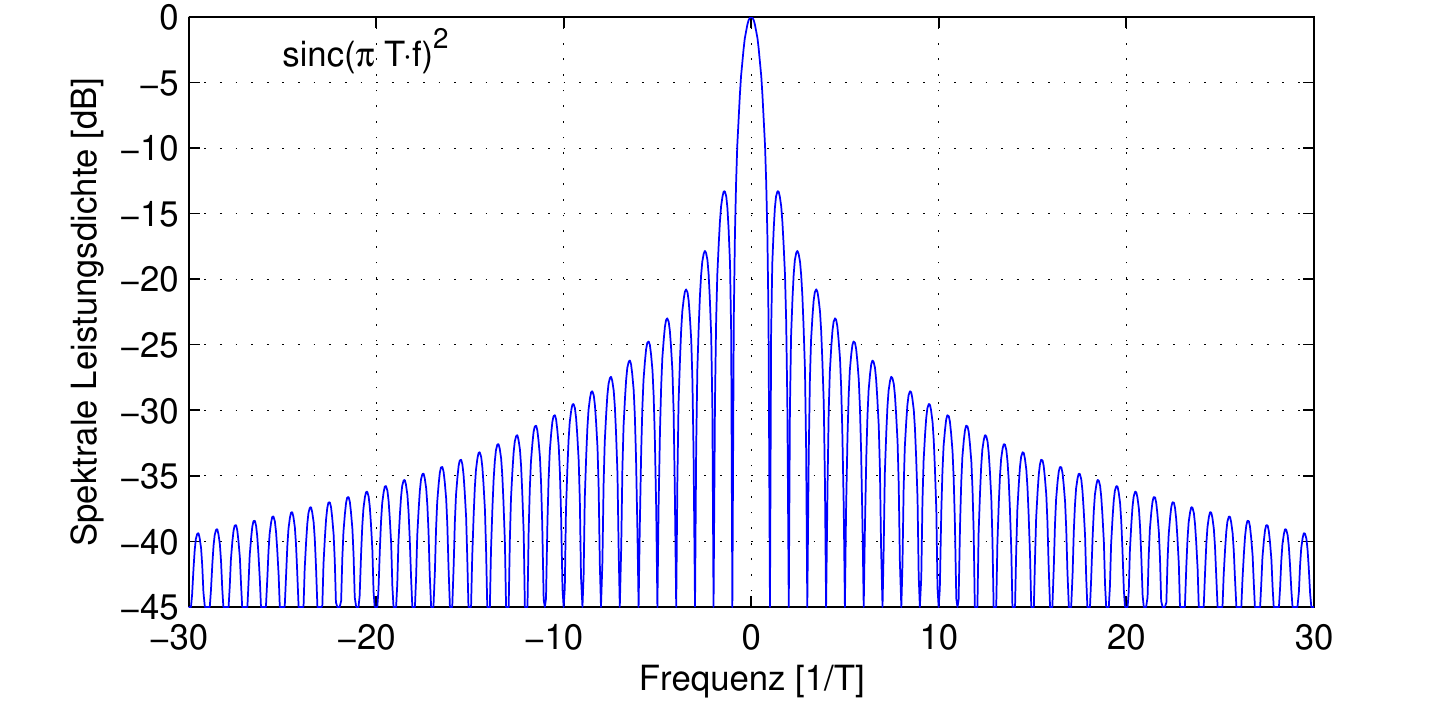}
	\caption{Bei zeitbegrenzten Signalen der Länge $T$ greift die DFT sinc-förmige Bereiche aus dem Frequenzspektrum heraus, deren Breite mit $1/T$ skaliert. Das Maximum der sinc-Funktion einer benachbarten Frequenz befindet sich genau an der ersten Nullstelle.}
	\label{fig:sinc}
\end{figure}
Der Leck-Effekt bezeichnet eine Verbreiterung von berechneten Frequenzspektren und tritt bei der Berechnung von Fouriertransformierten aus zeitlich begrenzten Signalen auf.
Die diskrete Fouriertransformation, wie sie hier Anwendung findet (Gl. \ref{eq:fourier}), arbeitet auf Signalabschnitten der Länge $T$, was einer Multiplikation des Gesamtsignals mit einer Rechteckfunktion $\mathrm{rect}(t/T)$ entspricht.
Nach dem Faltungstheorem führt diese Multiplikation im Zeitraum zu einer Faltung im Frequenzraum mit der entsprechenden Transformierten
\begin{equation}
\mathrm{FT}\bigl(\mathrm{rect}(t/T)\bigr)
= T\cdot \mathrm{sinc}(\pi T\cdot f)
\qquad\textnormal{mit}\qquad
\mathrm{sinc}(x)\equiv
\begin{cases}
  1, & x=0\\
  \sin(x)/x,  & x\neq0
\end{cases}
\end{equation}
Die Frequenzamplituden $\tilde{a}_i$ sind also genau genommen herausgegriffene Bereiche aus dem intrinsischen Spektrum $\tilde a(f)$:
\begin{equation}
\tilde{a}_i = \int_{-\infty}^{\infty}
\tilde a(f) \cdot T\,\mathrm{sinc}(\pi T\cdot (f-f_i))\,\mathrm{d}f
\label{eq:sincintegral}
\end{equation}
Die Leistungsdichte der $\mathrm{sinc}$-Funktion ist in Abb. \ref{fig:sinc} logarithmisch dargestellt.
Sie fällt nur mit $(T/\Delta f)^2$ ab, weshalb starke Signale sich auf große Frequenzbereiche auswirken können.

\subsubsection{Theoretisches Spektrum des Quantenrauschens}
\label{sec:theospec}
Welches Spektrum hat nun ein kohärentes Quantenrauschen?
Wie in Kap. \ref{kap:quantenstatistik} dargelegt treffen alle Photonen unabhängig zu beliebigen Zeiten ein.

Die Basisfunktionen der DFT ($e^{-2\pi\mathrm{i}\cdot\frac{jk}{N}}$, Gl. \ref{eq:fourier}) sind zueinander orthogonal und alle gleichermaßen normiert.
Deshalb beeinflusst jedes eintreffende Photon alle Fourierkomponenten gleich stark und unabhängig.
Das erwartete Spektrum der Detektorintensität entspricht also dem von weißem Rauschen.

Auch die $\mathrm{sinc}$-Funktionen der DFT (Gl. \ref{eq:sincintegral}) sind orthogonal zueinander.
Über ein weißes Spektrum gefaltet werden auch deren Amplituden unkorreliert sein.
Allerdings wirken sich Signale bei manchen Frequenzen im Spektrum gleichsinnig auf benachbarte sinc-Funktionen aus und gegensinnig auf andere.
Bei einem flachen Spektrum halten diese sich genau die Waage.
Wenn das Spektrum aber nach dem Detektorausgang nicht mehr komplett flach ist (Abb. \ref{fig:spectrum_spikes}), können dadurch Korrelationen entstehen.
Untersucht wird das in Kap.~\ref{sec:korr}.

\subsection{Entropie von Zufallsvariablen}
Die aus der Fouriertransformation erhaltenen Frequenzamplituden folgen einer Zufallsverteilung, sie enthalten deshalb "`Entropie"'.
Allerdings ist auch noch Systematik enthalten, und zur optimalen Weiterverarbeitung brauchen wir eine präzise Definition dieser Größen.

Informationstheoretische Entropie ist ein Maß für die Unsicherheit (Zufälligkeit) in einer Zufallsvariable $X$ mit möglichen Werten $x\in\mathcal{X}$ \cite{cover1991elements}.
Sie hängt nur von der Verteilungsfunktion der Variable (Ereigniswahrscheinlichkeiten $p(x)$) ab, aber nicht vom konkreten Wert der Variable $x$.

Je nach Anwendung existieren verschiedene Definitionen von Entropie.
Interessant für dieses Experiment sind die Shannonentropie
\begin{equation}
H_1(X) = \sum_{x\in\mathcal{X}} p(x)\cdot\log_2\frac{1}{p(x)}
\qquad\quad\textnormal{mit }0\log_2\frac{1}{0}:=0
\ ,
\label{eq:shannon}
\end{equation}
welche ein Maß für die durchschnittliche Zufälligkeit in $X$ ist und die Min-Entropie
\begin{equation}
H_\infty(X)
= \min_{x\in\mathcal{X}}\log_2\frac{1}{p(x)}
= \log_2\frac{1}{\max(p(x))}
\ ,
\label{eq:minentropie}
\end{equation}
die die minimale Zufälligkeit in $X$ beschreibt.
Der Index spezifiziert hier die Entropien als Spezialfall der Renyi-Entropie $H_\alpha$ \citep{renyi1961}, die hier nicht in allgemeiner Form benötigt wird.

Sobald eine zweite Zufallsvariable $Y$ bekannt ist, die Information über $X$ enthält, verringert sich die Entropie zur bedingten Entropie
\begin{equation}
H_1(X|Y)
= \sum_{y\in\mathcal{Y}} p(y)\sum_{x\in\mathcal{X}}p(x|y)\cdot\log_2\frac{1}{p(x|y)}
\label{eq:shannon_bed}
\end{equation}
bzw.
\begin{equation}
H_\infty(X|Y) = \min_{y\in\mathcal{Y}} \min_{x\in\mathcal{X}}\left(\log_2\frac{1}{p(x|y)}\right)
= \log_2\frac{1}{\max(p(x|y))}
\label{eq:min_bed}
\end{equation}
(Definition nach \cite{Renner2005}),
wobei $p(x|y)$ die bedingte Wahrscheinlichkeit ist, dass $x$ auftritt, sobald $y$ schon aufgetreten ist.
$H_1(X|Y)$ ist also ein Maß für die mittlere verbleibende Zufälligkeit in $X$ wenn $Y$ bekannt ist \citep{applebaum1996probability} und $H_\infty(X|Y)$ für die minimale verbleibende Zufälligkeit.

Eine mögliche Abschätzung für die bedingte Shannonentropie wurde im Vorgängerprojekt \cite{Gabriel2010} bereits gezeigt, wo das Gesamtsignal $T$ von $K$ und $Q$ abhängt.
Wird nämlich die Transinformation $I(T;K)$ betrachtet, so gilt nach \citep{cover1991elements} stets $H_1(T|K) = H_1(T) - I(T;K)$ und $I(T;K) \leq H_1(K)$ und somit
\begin{equation}
H_1(T|K) = H_1(T) - I(T;K) \geq H_1(T) - H_1(K)
\ .
\end{equation}
$H_1(T)$ und $H_1(K)$ wurden dort aus der Messung von Gesamt- und Elektronikrauschen abgeschätzt, und somit eine Mindestgrenze für die bedingte Entropie ermittelt.
Zur Steigerung der Datenrate wurde hier aber eine engere Grenze durch die direkte Berechnung nach Gl.~\ref{eq:shannon_bed} bzw. \ref{eq:min_bed} gesucht.

Im vorliegenden Fall (Abb.~\ref{fig:infoflow}) ist die Zufallsvariable $R$ bekannt, die eine Funktion des Quanten\-rauschens $Q$ und des (möglicherweise klassischen) Elektronik\-rauschens $K$ ist.
Weil $K$ nicht als Ergebnis eines verstandenen quantenmechanischen Zufallsprozesses beschrieben werden kann, soll der schlimmste Fall eines vollständig deterministischen $K$ angenommen werden.
Die Entropie, welche danach in $R$ noch übrig bleibt, ist also gegeben durch
\begin{equation}
H_1(R|K)
\qquad\textnormal{bzw.}\qquad
H_\infty(R|K)
\ .
\end{equation}

Welche von diesen Entropien nun verwendet wird, hängt von den weiteren Annahmen und Sicherheitsanforderungen ab.
Wenn der konkrete Verlauf von $K$ unbekannt bleibt und sehr viele Zufallszahlen zusammen gehasht werden, dann greift der Satz der Asymp\-to\-tischen Gleichverteilung \citep{cover1991elements} und die Entropie konvergiert zur Shannonentropie $H_1$.
Die Wahrscheinlichkeit einer deutlich geringeren Entropie als der Shannonentropie konvergiert für lange Zahlenfolgen gegen Null.

Anders der Fall, wenn ein Angreifer das $K$ genau kennt oder sogar kontrolliert.
In diesem Fall weiß er genau, wenn die Entropie verringert ist, und kann das nutzen, um Information über die Zufallszahlen zu erhalten.
In diesem Fall bleibt nur noch $H_{\infty}$ übrig.
Eine geschickte Art, die Zufallsbits aus den Amplituden zu berechnen wird sicherstellen, dass $H_\infty$ nicht viel geringer ausfällt als $H_1$.

\subsection{Extraktion des Zufalls}
\label{sec:extraktion}
Wie kann der vorhandene Zufall einer Variablen nun extrahiert werden, so dass keinerlei Systematik übrig bleibt?
Dass so eine Prozedur überhaupt möglich ist, wird durch das "`leftover hash lemma"' \cite{Impagliazzo1989leftover} gezeigt.
Benötigt wird hierzu eine Funktion die nicht effizient invertierbar ist.
Bei kryptographischen Hashingfunktionen ist das höchstwahrscheinlich der Fall, es soll aber angemerkt sein dass mathematische Beweise dafür noch ausstehen.
Existierende Algorithmen für diese Anwendung sind SHA512 oder Whirlpool \citep{barreto2000whirlpool}.

Das Hashingverhältnis, also die Menge an Zufallszahlen, die aus einer gegebenen Datenmenge erzeugt werden dürfen, folgt aus der bedingten Entropie.
In \cite{Renner2005} wurde daraus die extrahierbare Entropie berechnet: $H_{\textnormal{ext}}^{\varepsilon}$:
\begin{equation}
H_{\textnormal{ext}}^{\varepsilon}
\geq H_{\infty}(R|K) - 2\log_2(1/\varepsilon)
\end{equation}
Der Parameter $\varepsilon$ steht dabei für die maximal verbleibende Restwahrscheinlichkeit dass die resultierende Verteilung nicht exakt gleichverteilt ist und kann durch Anpassung des Hashingverhältnisses so klein wie gewünscht gewählt werden.
Um besonders viel Zufall extrahieren zu können, muss daher die Maximalwahrscheinlichkeit für ein Ereignis bei einem schlimmstmöglichen gegebenen klassischen Rauschen (Gl.~\ref{eq:min_bed}) gering gehalten werden, und viele Bits gemeinsam gehasht werden um Verluste durch den $\varepsilon$-Faktor klein zu halten.

\subsection{Binning}
\label{kap:binning}
Die weitere Verarbeitung der Amplituden wird für jede Frequenz getrennt durchgeführt.
Nach der DFT sind diese nach dem Zentralen Grenzwertsatz als gewichtete Summe sehr vieler Einzelwerte annähernd normalverteilt.
Wegen dieser nicht-flachen Verteilung und dem direkten Einfluss des Elektronik\-rauschens besitzen diese deutlich weniger bedingte Entropie als ihrer Beschreibungslänge entspräche.
Die Amplituden sollen nun so auf eine Folge von digitalen Zufallsbits abgebildet werden, so dass deren bedingte Entropie möglichst so groß wird wie ihre Beschreibungslänge, was einer perfekten Zufälligkeit entspricht.
Als eine mögliche Methode wird über die Verteilungsfunktion ein Binning gelegt und jeder Amplitude der digitale Wert des Bins zugewiesen in den sie fällt.

Weil das Quantenrauschen unabhängig vom Elektronikrauschen ist, addieren sich die Varianzen $\sigma^2$ der Beiden
\begin{equation}
\sigma_T^2 = \sigma_K^2 + \sigma_Q^2
\ .
\end{equation}
Das Quantenrauschen ist bereits im Detektor normalverteilt (Kap.~\ref{sec:theospec}) und bleibt es auch nach der DFT.
Die nicht direkt messbare Verteilung des Quantenrauschens ist somit bekannt als Normalverteilung mit Varianz $\sigma_T^2 - \sigma_K^2$.

Hierdurch ist auch die bedingte Entropie (Gl.~\ref{eq:shannon_bed},~\ref{eq:min_bed}) berechenbar.
Summen, Minima und Maxima laufen dann über alle kontinuierlichen Werte von $K$ und über alle Bins, deren Inhalt von der um $k$ verschobenen $Q$-Verteilung gegeben ist.

\FloatBarrier
\subsubsection{Empirisch flächengleiches Binning}
\begin{figure}
	\centering
		\includegraphics[width=\textwidth]{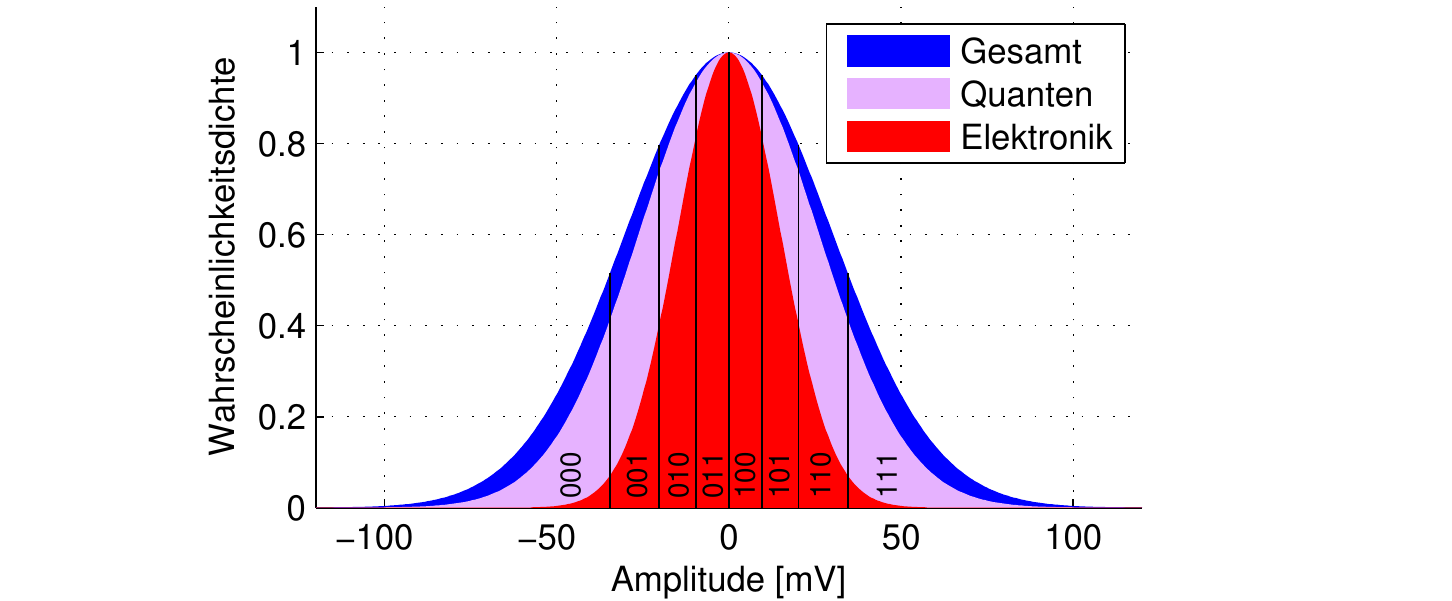}
	\caption{Empirisch flächengleiches Binning: Die Gesamtverteilung (blau) wird in Bins gleicher Fläche aufgeteilt. Die einzelnen Verteilungen sind hier übereinander gezeichnet und auf ihr Maximum normiert. Resultierende Zufalls-Bitsequenzen laufen von 0 bis $2^n-1$ (dargestellt ist $n=3$).}
	\label{fig:binning_equalarea}
\end{figure}

Der erste und naheliegende Ansatz eines solchen Binnings geht auf das Vorgängerprojekt dieser Arbeit \cite{Gabriel2010} zurück.
Die Verteilungsfunktion wird durch ein Ensemble vieler ($m$) Messwerte empirisch ermittelt und dann in $2^n$ Bins gleicher Fläche eingeteilt (Abb.~\ref{fig:binning_equalarea}).
Die Gleichverteilung ist damit bereits sichergestellt.

Der Entropieverlust durch Kenntnis von $K$ entspricht der Information in $K$ über $R$.
Diese Information ist im Mittel gering, denn $K$ gibt nur eine Tendenz der Verteilung vor, worauf das Quantenrauschen $Q$ immer noch alle Möglichkeiten zur Verfügung hat.

Zur Abschätzung der Entropie wir die berechtigte Annahme normalverteilter Größen $\rho(x)=\exp(-\frac12(x/\sigma)^2)/(\sqrt{2\pi}\sigma)$ und die Näherung vieler kleiner Bins mit annähernd homogener Wahrscheinlichkeitsdichte innerhalb jedes Bins gemacht.
Die Binbreite $b$ ist in diesem Fall
\begin{equation}
b(x) = \frac{1}{2^n\rho_T(x)}
\qquad\textnormal{wegen}\quad
b\cdot\rho_T=\frac{1}{2^n}
\ .
\end{equation}
Die Entropie kann nun anstatt einer Summe als Integral genähert werden:
\begin{equation}
H_1(X) = \sum_{x\in\mathcal{X}} p(x)\cdot\log_2\frac{1}{p(x)}
\quad\rightarrow\quad
\int_{-\infty}^{\infty}
\rho(x)\cdot\log_2\frac{1}{\rho(x)\cdot b(x)}\,\dif x
\end{equation}
Für gegebenes klassisches Rauschen $K=k$ folgt daraus die Entropie
\begin{equation}
H_1(R|K=k) = 
\int_{-\infty}^{\infty}
\rho_Q(x-k)\cdot\log_2\frac{2^n\cdot \rho_T(x)}{\rho_Q(x-k)}\,\dif x =
\end{equation}
\begin{equation}
= n - \log_2\left(\frac{\sigma_T}{\sigma_Q}\exp\left(-\frac12\left(\frac{\sigma_k^2-k^2}{\sigma_T^2}\right)\right)\right)
\end{equation}
und somit die bedingte Entropie
\begin{equation}
H_1(R|K) = 
\int_{-\infty}^{\infty}
\rho_K(k)\cdot H_1(R|K=k)\,\dif k
= n - \log_2\frac{\sigma_T}{\sigma_Q}
\end{equation}
Das bedeutet, der Verlust an Shannonentropie $I(R;K) = H_1(R) - H_1(R|K) = \log_2\frac{\sigma_T}{\sigma_Q}$ ist nicht von der Binanzahl $2^n$ abhängig, sondern (bei nicht zu großen Bins) nur vom Signal-Rausch-Verhältnis.
Im vorliegenden Fall von $\sigma_T^2/\sigma_K^2\approx 10\unit{dB}$ ist der Entropieverlust gerade mal $0{,}076$ Bits pro Sample.

Weitere Entropieverluste folgen durch langsame Schwankungen der Intensität von $\Delta\sigma/\sigma<2\%$ zu $\Delta H_1=(\Delta\sigma/\sigma)^2/\ln2=6\ee{-4}$ pro Sample.
Dadurch dass die Verteilung nach $m\approx5\ee5$ Samples in Bins gleicher Samplezahl gezwungen wird werden die theoretischen Möglichkeiten reduziert und es folgt ein weiterer Entropieverlust von $\Delta H_1=\frac12\frac{n}{m}\log_2\left(\frac{2\pi}{e}\frac{m}{n}\right)=3\ee{-3}$ bei $n=8$ pro Sample.
Insgesamt wird die Entropie mit
\begin{equation}
H_1(R|K) \approx n - 0{,}08
\end{equation}
kaum verringert.

Anders sieht die Sache für die bedingte Min-Entropie aus.
Diese wird umso kleiner je weiter $k$ außen am Rand liegt, weil dort $Q$ effektiv weniger Bins zur Verfügung hat.
Für eine normalverteilte Größe wie im vorliegenden Fall kann $k$ prinzipiell beliebig weit außen liegen so dass die Entropie gegen Null konvergiert.
Auch wenn der Fall praktisch sehr unwahrscheinlich ist, lässt sich dennoch keine unbedingte Sicherheit mehr beweisen.

\FloatBarrier
\subsubsection{Zyklisch äquidistantes Binning}
\begin{figure}[bt]
	\centering
		\includegraphics[width=\textwidth]{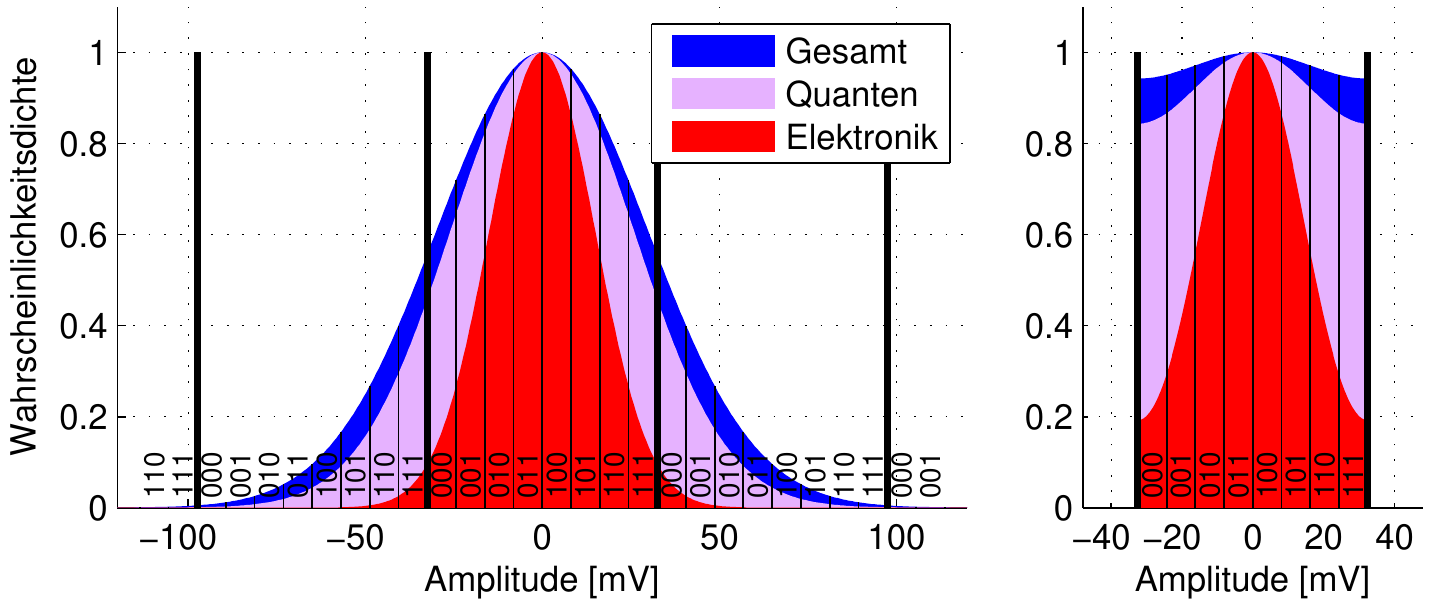}
	\caption{Zyklisch äquidistantes Binning: Die Gesamtverteilung wird in gleichbreite Bins aufgeteilt (links). Die zugeordneten Bitsquenzen wiederholen sich nach einer Breite $B$ (hier $B=2{,}5\,\sigma_Q$).
Effektiv ist die Wahrscheinlichkeit für jede Bitsequenz die Summe der entsprechenden Abschnitte (rechts). Alle Verteilungen sind zur Darstellung auf ihr Maximum normiert.}
	\label{fig:binning_wrap}
\end{figure}

Um $H_{\infty}$ groß zu halten, wurde ein Binning gewählt, bei dem die Kenntnis von $K$ auch im schlimmsten Fall nicht viel über das Ergebnis verraten kann.
Die Verteilung wird dabei in $2^n$ gleichbreite Bins der Breite $b$ (Abb.~\ref{fig:binning_wrap} links) eingeteilt, wobei sich die zugewiesenen Binärwerte nach $B=2^n\cdot b$ wiederholen.
Hierdurch ist die Wahrscheinlichkeit einzelner Binärwerte vorerst nicht gleichverteilt (Abb.~\ref{fig:binning_wrap} rechts).
Für kleinere $B$ nimmt die Gleichmäßigkeit der Verteilung aber zu, so dass das Wahrscheinlichkeitsverhältnis $\max(p)/\langle p\rangle$ bei $B=\sigma$ immerhin $1+5\ee{-9}$ und bei $B=0{,}5\,\sigma$ bereits $1+1\ee{-34}$ beträgt.
Das kommt einer Gleichverteilung so nahe dass die Abweichung kein limitierender Faktor mehr ist.

Hierfür soll nun die bedingte Entropie ermittelt werden.
Bei gegebenem $k$ folgt das Signal wieder der (um $k$ verschobenen) Quantenrauschverteilung (violett in Abb.~\ref{fig:binning_wrap} rechts).
Die Verteilung ändert aber bei Verschiebung ihre Eigenschaften nicht, weil die Binbreiten überall gleich sind.
Die bedingte Entropie folgt also aus der Form dieser Verteilung.
Für den Fall einer sehr flachen Verteilung, also wenn $B\lessapprox \sigma_Q$, nähern sich wegen $p(r|k)\rightarrow2^{-n}$ sowohl bedingte Shannon- als auch bedingte Min-Entropie an die Beschreibungslänge der Bitsequenz an, das heißt, durch Kenntnis von $K$ gewinnt man keinerlei Information über das Ergebnis.

Spätestens hier stellt sich die Frage, wie fein das Binning eigentlich gemacht werden darf, bzw. wodurch die Anzahl der extrahierbaren Bits pro Sample in diesem Schema limitiert ist.
Klar ist, dass keinesfalls mehr Bits extrahiert werden dürfen als die vom Oszilloskop empfangene Datenmenge, weil spätestens dann versteckte Abhängigkeiten in den Daten auftreten müssen.
Tatsächlich sind die Eingangswerte $T$ in die DFT diskrete 8-Bit Werte und daher sind auch die Frequenzamplituden $R$ diskret.
Bei sehr kleinen Binbreiten würden die Werte deshalb immer in bestimmte Bins fallen und in andere nicht.
Die mittlere und Maximalwahrscheinlichkeit pro Bin ist damit nach unten beschränkt, und damit die Entropie nach oben.

Die genaue Bestimmung der Wahrscheinlichkeiten einzelner Bins unter Berücksichtigung versteckter Abhängigkeiten geht über den Rahmen dieser Arbeit hinaus.
Auch empirische Abschätzungen sind schwierig, weil für eine ausreichende Statistik deutlich mehr DFTs durchgeführt werden müssen als die Binanzahl $2^n$.
Für $n=16$ ergab eine solche Simulation, dass die Wahrscheinlichkeit in keinem Bin signifikant größer war als die simulationsbedingten statistischen Fluktuationen von $2{,}5\cdot\langle p\rangle$.

\subsection{Zufallstests}
Zur empirischen Überprüfung, ob das Ergebnis nun richtige Zufallszahlen sind oder nicht werden gerne Zufallstests hergenommen wie z.\,B. die "`Diehard"' Sammlung \citep{marsaglia1996diehard}.
Allerdings kann bei Zufallszahlen per Definition jede Folge erscheinen, so dass aus keinem Ergebnis strikt eine Nicht-Zufälligkeit geschlossen werden kann.
Vorhandene Tests können also nur darauf hinweisen, wenn bestimmte Systematiken vorliegen, die sehr unwahrscheinlich auftauchen sollten.

Die Aussagekraft solcher Tests ist beschränkt.
Wenn ein Test negativ ausfällt, sind die Zahlen wahrscheinlich nicht zufällig.
Wird er aber bestanden, so heißt das keineswegs dass die Zahlen tatsächlich zufällig sind.
Weiterhin kann kaum getestet werden, wie zufällig die Zahlen noch sind nachdem Einflüsse wie das Elektronikrauschen bekannt werden würden, weil das Programm diese Einflüsse dazu verstehen müsste.
Die hier erzeugten Daten werden (nach Abschluss dieser Arbeit) solchen Tests unterzogen werden, doch ersetzt das keine sorgfältige Analyse der Erzeugungsalgorithmen.

\newpage
\FloatBarrier
\section{Experimenteller Aufbau}
\begin{figure}
	\centering
		\includegraphics[width=\textwidth]{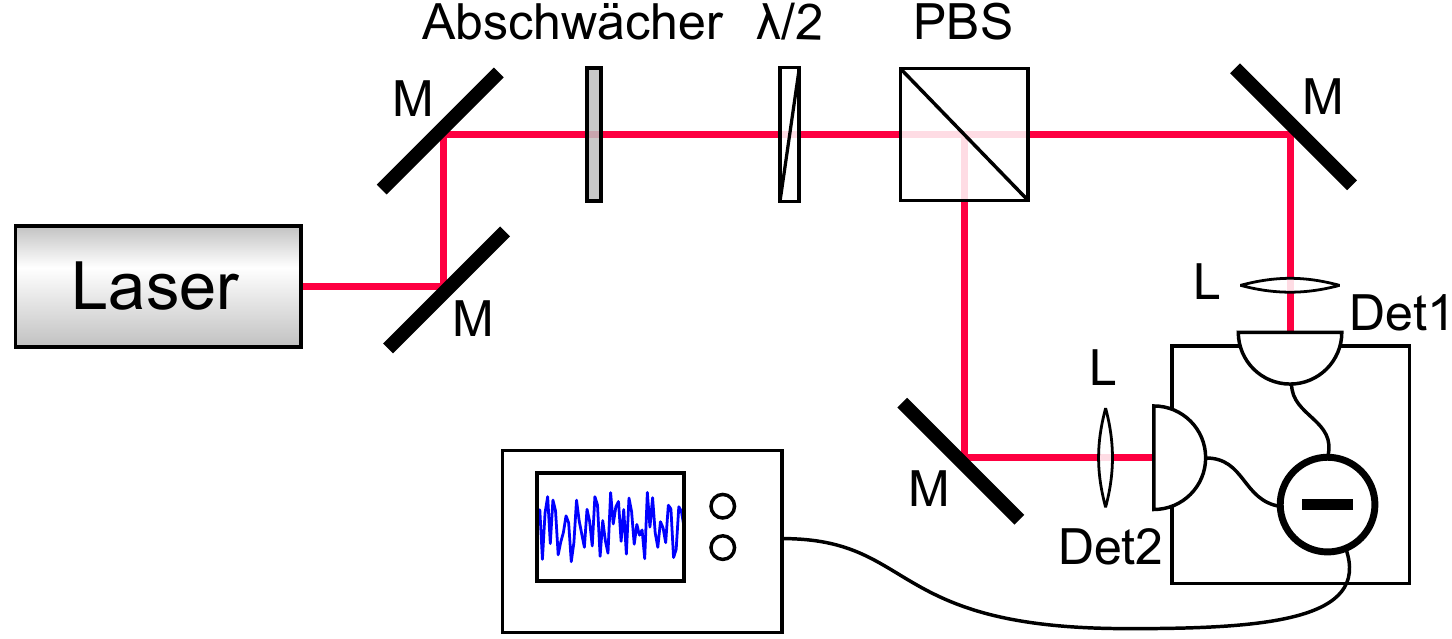}
	\caption{Verwendeter Messaufbau: Laser, Justierspiegel M, variabler Abschwächer, $\lambda/2$-Platte zum justieren der Polarisationsrichtung, Polarisierender Strahlteiler (PBS), Umlenkspiegel M, Fokussierlinsen L, Detektoren, Subtraktionselektronik und Ausleseoszilloskop.}
	\label{aufbau_skizze}
\end{figure}
\begin{figure}[t!]
	\centering
		\includegraphics[width=0.82\textwidth]{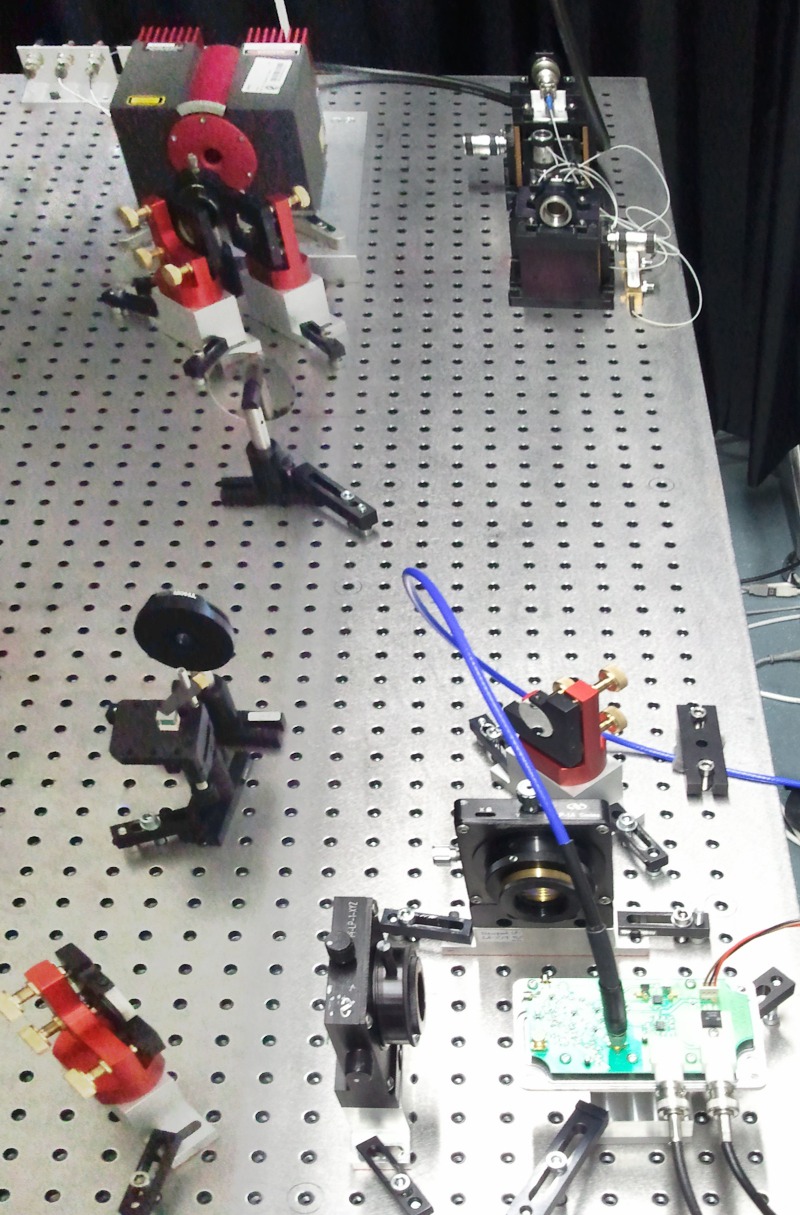}
    \caption{Kompletter Messaufbau von Laser (links oben) bis Detektor (rechts unten). Der Modulator (rechts oben) inklusive Fasern und Einkopplungstischen ist hier nicht in den Strahlengang integriert. Der Laserstrahl ist unsichtbar.}
	\label{aufbau_foto}
\end{figure}

Der experimentelle Aufbau des Zufallsgenerators (Abb. \ref{aufbau_skizze} und \ref{aufbau_foto}) wurde so einfach wie möglich gehalten und lässt die Struktur zur Homodyndetektion (Kapitel \ref{sec:homodyn}) noch direkt erkennen.
Als Lichtquelle diente ein rauscharmer Infrarotlaser\footnote{\href{http://innolight.de/fileadmin/user_upload/produktblatt_pdf/Produktblatt_Mephisto.pdf}{Mephisto von InnoLight}} mit $1064\unit{nm}$ Ausgangswellenlänge.
Ein konstanter Strahlabschwächer reduzierte die Strahlleistung auf $10\unit{mW}$.
Über zwei $90\degr$-Spiegel wurden die Positions- und Richtungsfreiheitsgrade des Strahls ausgerichtet.
Darauf folgte ein variabler Strahlabschwächer in Form einer Drehscheibe mit Durchlässigkeits\-gradient, um die Leistung flexibel anpassen zu können.

\begin{figure}[t]
	\centering
	\begin{minipage}[b]{\textwidth}
	\includegraphics[width=0.47\textwidth]{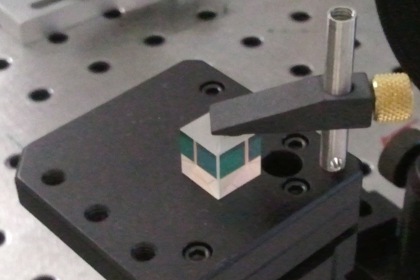}
	\hfill
	\includegraphics[width=0.47\textwidth]{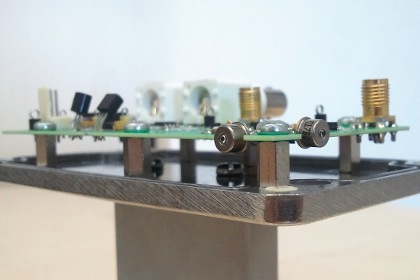}
	\end{minipage}
	\caption{Links: Polarisierender Strahlteiler. Rechts: Detektor mit den zwei Fotodioden im Vordergrund.}
	\label{fig:bs_detector}
\end{figure}

Der Strahlteiler zur Homodyndetektion wurde durch einen polarisierenden Strahlteiler (PBS) realisiert (Abb.~\ref{fig:bs_detector} links).
Hierzu wurde die lineare Polarisation des Laserstrahls mit einer drehbaren $\lambda/2$-Wellenplatte auf $45\degr$ eingestellt und im Strahlteiler auf gleichintensive Stahlen orthogonaler Polarisation aufgeteilt.
Somit blieb das Leistungsverhältnis beider Teilstrahlen auf eine Genauigkeit von ca. $0{,}5\%$ justierbar.

Hinter dem PBS wurden beide Strahlen mit Hilfe von Spiegeln auf die Detektoren gelenkt, die sich auf einer gemeinsamen Platine befanden.
Um den auf mehrere Millimeter ausgedehnten Laserstrahl auf die kleine Detektorfläche zu bündeln wurden noch entsprechende Sammellinsen ($f=50\unit{mm}$) vor den Detektoren positioniert.

Für handelsübliche Detektoren ist es sehr schwierig, Leistungsschwankungen im GHz-Bereich  zu detektieren, weil dann kleinste Kapazitäten und Induktivitäten der Messelektronik bedeutsam werden.
Deshalb kam hier ein Detektor der University of New South Wales zum Einsatz (Abb.~\ref{fig:bs_detector} rechts), der speziell für den Frequenzbereich bis $1\unit{GHz}$ entwickelt wurde.
Bei diesem wird die Signalsubtraktion direkt auf der Platine durchgeführt.
Unvermeidbar ist bei so einem empfindlichen Detektor allerdings der unerwünschte Einfang von Funksignalen verschiedenster Art, was die Analyse (Kap.~\ref{sec:funk}) deutlich zeigt.

Aufgezeichnet wurde das Signal mit einem 8-Bit Digitaloszilloskop\footnote{\href{http://cdn.lecroy.com/files/pdf/lecroy_waverunner_6_zi_datasheet.pdf}{LeCroy WaveRunner 640Zi}} mit einer analogen Bandbreite von $4\unit{GHz}$ und einer maximalen Abtastrate von $20\unit{GS/s}$.
Das Oszilloskop konnte alle zwei Sekunden eine Binärdatei mit $2\ee{7}$ Messpunkten abspeichern.
Das würde die Datenrate für eine Echtzeit-Anwendung ausbremsen, reicht aber zur Demonstration des Prinzips und kann in einer späteren Anwendung umgangen werden.

\FloatBarrier
\section{Messungen}
\label{sec:messung}
\subsection{Datenaufnahme}
\begin{figure}[bth]
	\centering
		\includegraphics[width=\textwidth]{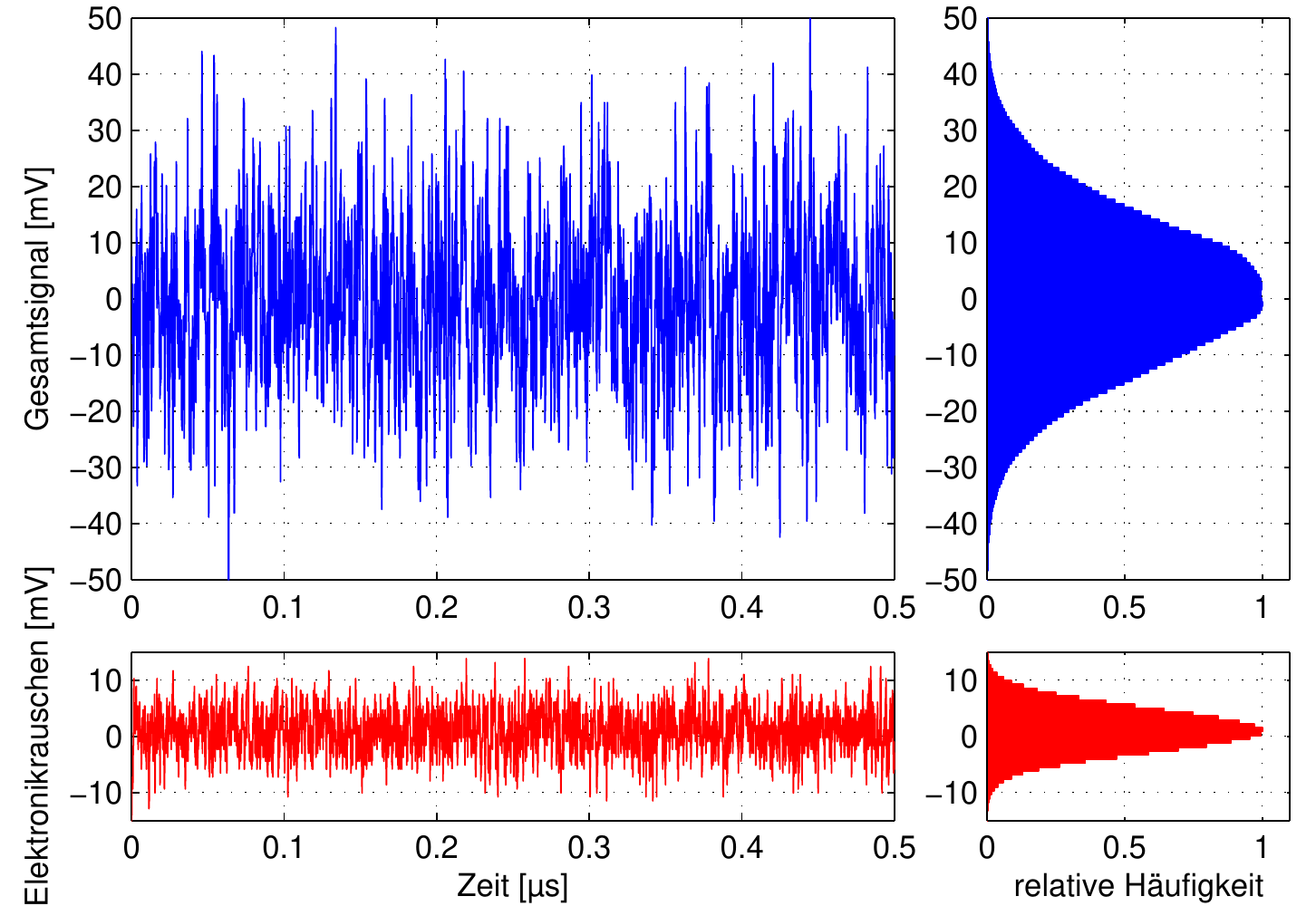}
	\caption{Links: Zeitlicher Verlauf des Detektorsignals zu einem beliebigen Zeitpunkt.
Rechts: Verteilungsfunktion der 8-Bit Oszilloskopdaten.
Oben: Signal mit Quantenrauschen.
Unten: Nur Elektronikrauschen.}
	\label{fig:zeitreihe}
\end{figure}

Im Normalbetrieb wird die Laserleistung konstant gehalten und die $\lambda/2$-Platte einmal so eingestellt, dass beide Detektoren ein gleichstarkes Ausgangssignal liefern.
Das Oszilloskop zeichnet das Differenzsignal auf.
Die Daten werden auf eine externe Festplatte gespeichert.
Auf diese Weise wurden über ein Wochenende $6\unit{TB}$ an Rohdaten zur Analyse generiert.
Die Weiterverarbeitung der Daten am PC wird im Nachhinein vorgenommen.
Abb.~\ref{fig:zeitreihe} zeigt einen kurzen Zeitabschnitt des Signals und die Amplitudenverteilung mit und ohne Laserstrahl.

\FloatBarrier
\subsection{Das Spektrum}
\begin{figure}[bht]
	\centering
		\includegraphics[width=\textwidth]{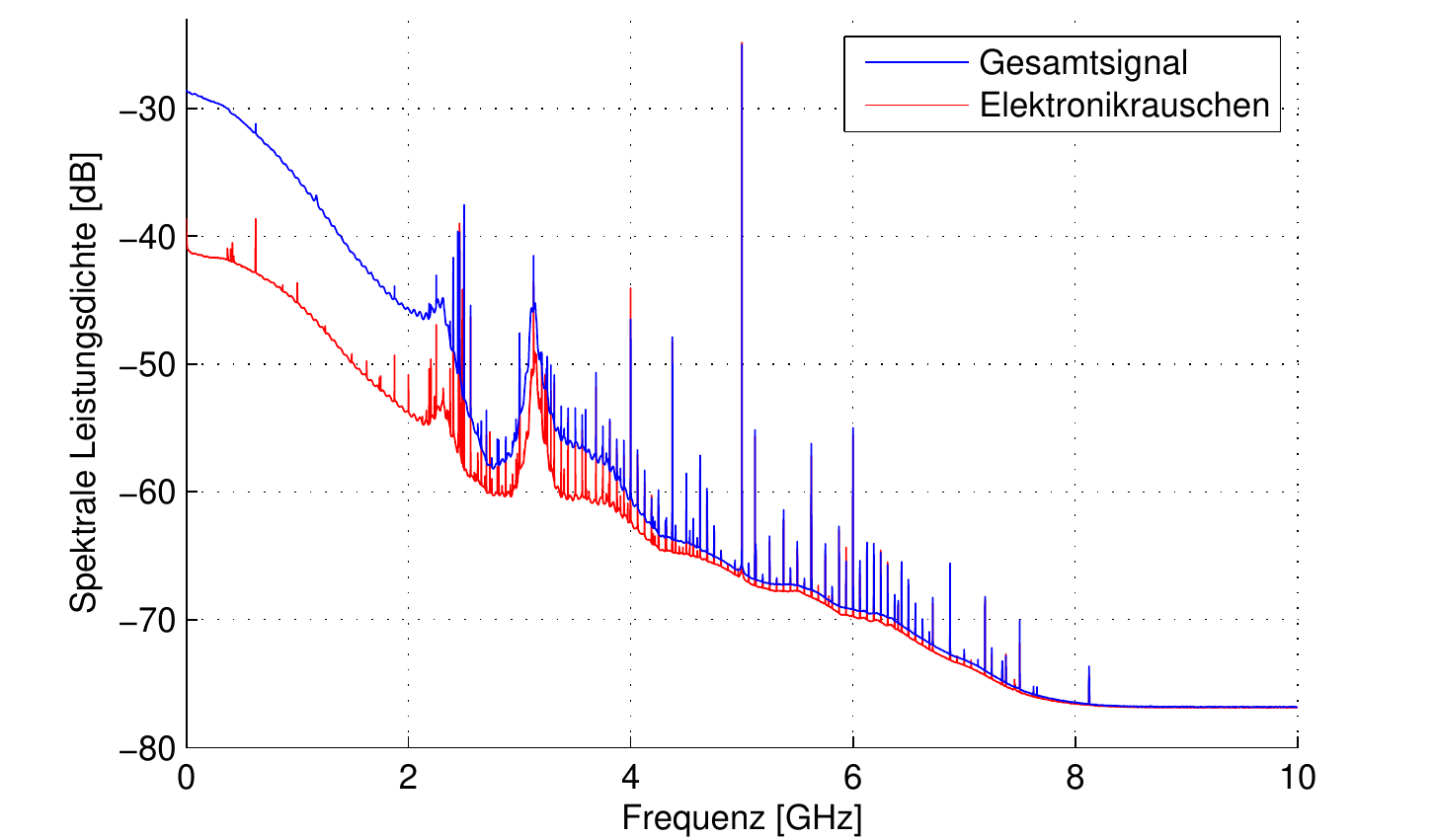}
	\caption{Spektrum des Detektorsignals mit und ohne Lasersignal. Es erscheinen zahlreiche scharf begrenzte Spitzen, besonders bei Bruchteilen der Samplingfrequenz $20\unit{GS/s}$.}
	\label{fig:spectrum_spikes}
\end{figure}
Das Fourierspektrum des Signals ist in Abb.~\ref{fig:spectrum_spikes} mit einer Dezibelskala dargestellt.
Für hohe Frequenzen fällt eine starke Abnahme des Signals um mehrere Größenordnungen auf.
Das liegt daran, dass Oszilloskop und Detektor für hohe Frequenzen als Tiefpass wirken.
Schließlich ist das Oszilloskop für $4\unit{GHz}$ und der Detektor nur für $1\unit{GHz}$ ausgelegt, was für aktuelle Maßstäbe bereits viel ist.
Die spezifische Form des Spektrums mit dem Hügel bei $3\unit{GHz}$ ist eine Charakteristik des Detektors, der nicht dafür ausgelegt ist, weit über $1\unit{GHz}$ das Signal getreu wiederzugeben.

Außerdem erscheinen zahlreiche scharf begrenzte Spitzen, die bei ganzzahligen Verhältnissen der Samplingfrequenz $20\unit{GS/s}$ besonders dominant sind.
Diese behalten bei Änderung der Fourierauflösung ihre Fläche, nicht aber ihre Höhe bei und sind Artefakte, die beim Digitalisierungsvorgang im Oszilloskop entstehen.
Dort wird die hohe Zeitauflösung durch Zusammenschaltung verschiedener AD-Wandler erreicht, die nicht exakt gleich sind und beim Übergang Störsignale einfügen.
Die Spitzen sind für das Experiment relativ ungefährlich, weil sie einfach als klassischer Einfluss mitbehandelt werden können.
Trotzdem sind sie zur Kontrolle über das System und für eine hohe Datenrate nicht wünschenswert.
Aus diesem Grund wurden die Daten nicht mit der minimal nötigen Rate von $2\unit{GS/s}$ (Nyquist) sondern mit $20\unit{GS/s}$ aufgezeichnet, um den Einflussbereich zu hohen Frequenzen hin zu verschieben.

Im Nutzbereich bis $1\unit{GHz}$ zeigt sich bei $5\unit{mW}$ Strahlleistung ein Signal-Rausch-Abstand von etwa $10\unit{dB}$, also ein Varianzverhältnis von 10 bzw. ein Amplitudenverhältnis von 3.
Ein solch signifikanter Rauschanteil würde die Information, die man über das Quantenrauschen erhalten kann, stark limitieren.
In der Anwendung für Zufallszahlen muss aber auch nicht auf das Quantenrauschen zurückgeschlossen werden können.
Vielmehr reicht es aus, dass aus dem klassischen Rauschen nicht viel Information über das Ergebnis hervorgeht.
Deshalb führt der elektronische Rauschanteil hier zu keiner drastischen Einschränkung.

\FloatBarrier
\subsection{Nachweis der Linearität (Abschwächemessung)}
\label{sec:linear}
\begin{figure}[thb]
	\centering
		\includegraphics[width=0.98\textwidth]{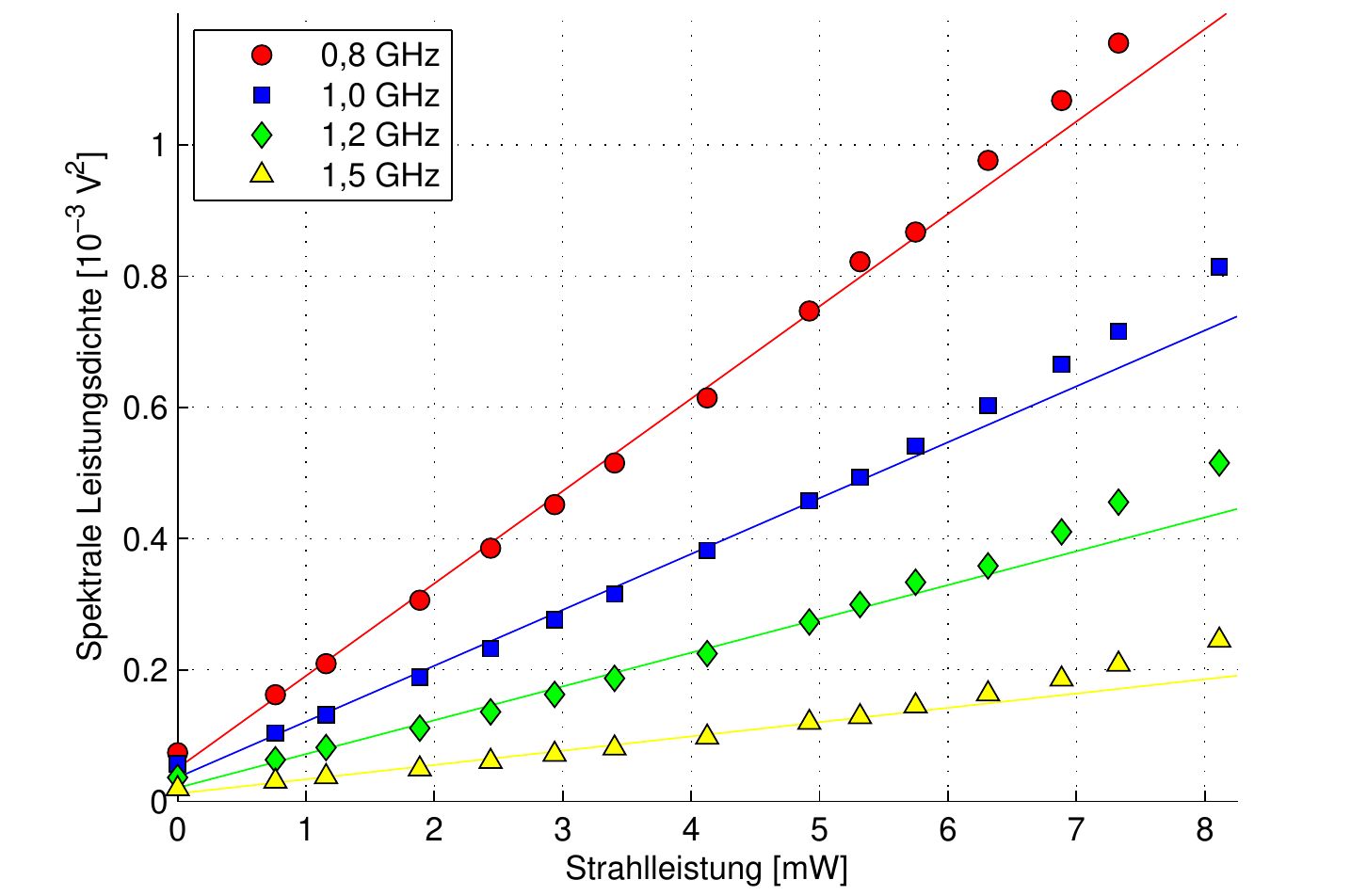}
	\caption{Spektralen Leistungsdichte in Abhängigkeit von der Strahlleistung.
Dargestellt sind vier ausgewählte Frequenzen aus dem Spektrum und jeweils lineare Fits für die Daten bis $6\unit{mW}$.}
	\label{fig:attenuation_plot}
\end{figure}
Wie in Kapitel \ref{kap:quantenstatistik} gezeigt, muss durch das Quantenrauschen die spektrale Leistungsdichte im Detektor linear von der Intensität des Eingangssignal abhängen.
Wäre dies nicht der Fall, so könnten andere unerwünschte Rauschquellen vorliegen, oder aber der Detektor wäre nicht linear, was zum Übersprechen zwischen unterschiedlichen Frequenzen führen würde.
Deshalb wurde untersucht, in welchem Leistungs- und Frequenzbereich der Detektor lineares Verhalten zeigt.
Hierfür wurde die Strahlleistung jeweils mit dem drehbaren Abschwächerrad variiert und die Leistung direkt im Strahl gemessen.
Danach nahm das Oszilloskop das Detektorsignal auf.
Das Messergebnis ist in Abb. \ref{fig:attenuation_plot} dargestellt und zeigt, dass die Linearität bei niedriger Strahlleistung und geringen Frequenzen bis ca. $6\unit{mW}$ bzw. $1{,}2\unit{GHz}$ sehr gut ist.
Stärkere Nichtlinearitäten traten erst über $8\unit{mW}$ und $1{,}5\unit{GHz}$ auf.

Für ein additives Rauschsignal, das unabhängig vom Quantenrauschen ist, wird wegen $\sigma^2(R) \sim \sigma^2(S) = \sigma^2(Q) + \sigma^2(K)$ ein konstanter Offset erwartet wie in Abb. \ref{fig:attenuation_plot} zu sehen, aber keinerlei anderer Einfluss.

\FloatBarrier
\subsection{Nachweis der Unabhängigkeit (Modulationsmessung)}
Die Unabhängigkeit verschiedener Frequenzkomponenten und damit der Quellen der Zufallszahlen kann noch genauer untersucht werden, indem der Effekt einer bekannten Signalanregung auf das Detektorsignal gemessen wird.
Im Idealfall werden am Detektor keine anderen Frequenzen angeregt als die des Eingangssignals.

Das lässt sich am deutlichsten durch einfache aufmodulierte Sinusschwingungen im Laserstrahl untersuchen.
Hierzu wurde ein elektrooptischer Modulator\footnote{\href{http://www.jdsu.com/ProductLiterature/apemicroanamod_ds_cc_ae_021306.pdf}{Modell 10020465 von JDSU}} an einen Funktionsgenerator\footnote{\href{http://www.sglabs.it/public/HP_837xxA\%20series.pdf}{Modell HP 83712A von Agilent}} angeschlossen und mit Hilfe von optischen Fasern (Abbildung \ref{aufbau_foto}, rechts oben) in den Strahlengang eingebaut.

\begin{figure}
	\centering
\includegraphics[width=1\textwidth]{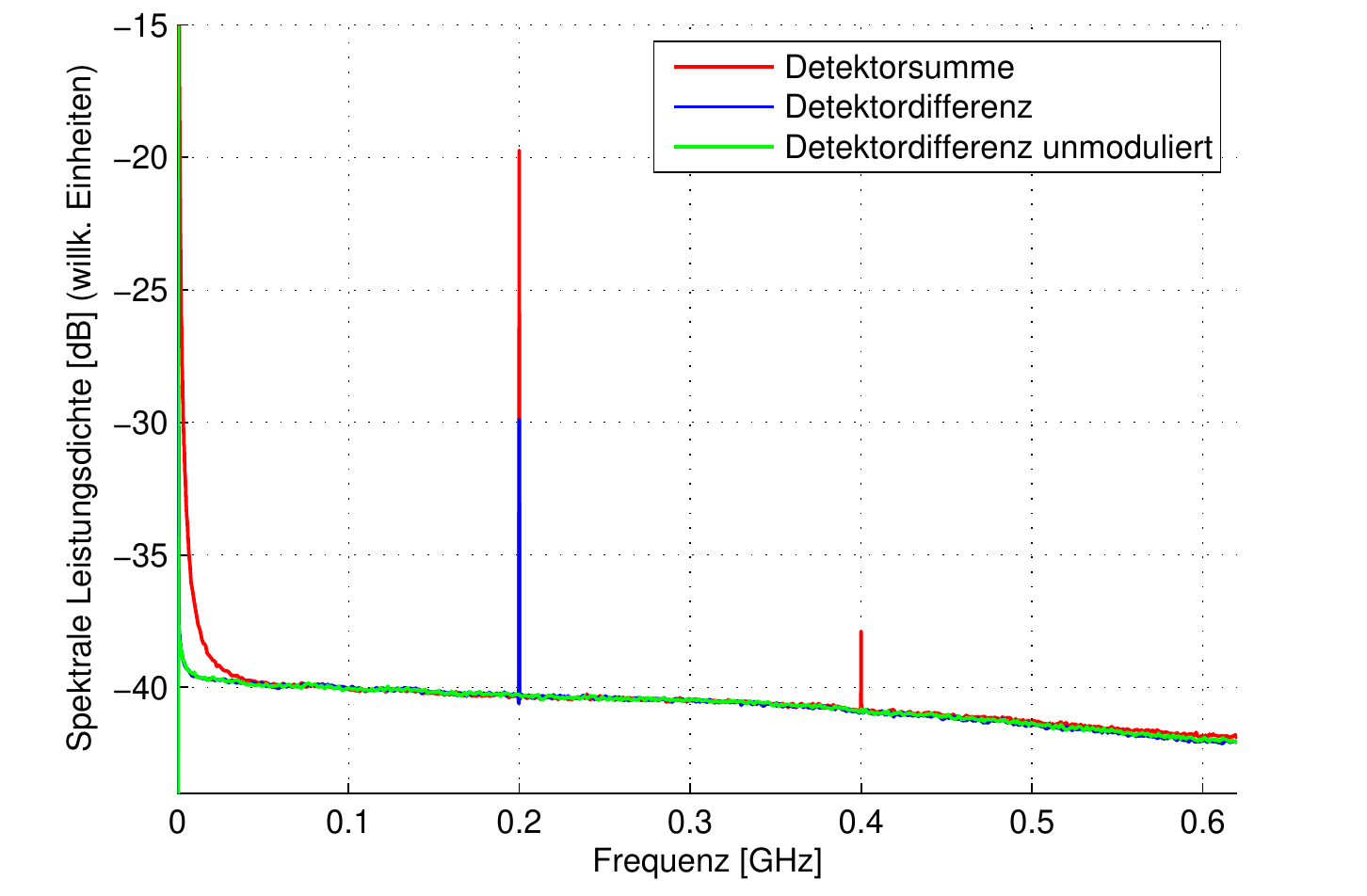}
	\caption{Spektrum des modulierten Lichtsignals. Bei $0{,}2\unit{GHz}$ wurde der Laserstrahl mit einem Signalgenerator moduliert. Bei $0{,}4\unit{GHz}$ erscheint die zweite Harmonische um $18\unit{dB}$ schwächer. Im Differenzsignal ist die Spitze um $10\unit{dB}$ unterdrückt.}
	\label{fig:spectrum_modulation}
\end{figure}

Mit diesem Aufbau wurde der Laserstrahl nacheinander von $100\unit{MHz}$ bis $1\unit{GHz}$ in $100\unit{MHz}$ Schritten moduliert.
Die Linienbreite von Signalgenerator und Detektorsignal war jeweils so schmal, dass mit unserer maximalen Auflösung von $1\unit{kHz}$ keine Verbreiterung festgestellt werden konnte.
Allerdings, wie in Abb. \ref{fig:spectrum_modulation} dargestellt, erschien das Signal der zweiten Harmonischen bei der doppelten Anregungsfrequenz.
Dieser Effekt lässt sich durch die Deformation der ursprünglichen Sinuswelle erklären, so dass weitere Komponenten in der Fourierreihe entstehen.
Alle höheren Harmonischen des Signalgenerators selbst befanden sich mindestens $48\unit{dB}$ unter dem Hauptsignal.
Im Detektorsignal dagegen erscheint die zweite Harmonische um $18\unit{dB}$ unter dem Hauptsignal, was akzeptabel klein ist, da sie bei der Messung der Zufallszahlen mit einem maximalen Signal-zu-Rauschabstand von $12\unit{dB}$ unter dem Elektronikrauschen begraben sein wird.

Der Plot zeigt außerdem dass diese Art von klassischem Rauschen im Minussignal um $10\unit{dB}$ unterdrückt ist.
Bei präziser Ausrichtung der $\lambda/2$-Platte, wie zur Aufnahme der Zufallszahlen, ist die Auslöschungung sogar noch deutlich stärker.

\FloatBarrier
\subsection{Untersuchung von Korrelationen}
\label{sec:korr}
Die gezeigte Linearität des Detektors und das geringe Übersprechen einzelner Frequenzen zu anderen ließ keine starken statistischen Abhängigkeiten verschiedener Frequenzamplituden erwarten.
Ein weiterer Test war die Messung von Korrelationen.

Im folgenden wurden Korrelationen zwischen jeweils zwei Komponenten
\begin{equation}
c(x, y) =
\frac{\langle \Delta x\cdot \Delta y\rangle}
{\sqrt{\langle \Delta x^2\rangle \cdot \langle \Delta y^2\rangle}}
\qquad\textnormal{mit}\qquad
\Delta x = x - \langle x\rangle
\end{equation}
anhand der Messdaten abgeschätzt.
Bei statistischer Unabhängigkeit sollten die Werte nahe bei Null liegen.
Berechnet wurden die Korrelationen zwischen Real- und Imaginärteil einzelner Frequenzkomponenten, die Korrelation zwischen benachbarten Frequenzen und die Korrelation aufeinanderfolgender Werte der selben Frequenz.
\begin{figure}[b!]
	\centering
		\includegraphics[width=\textwidth]{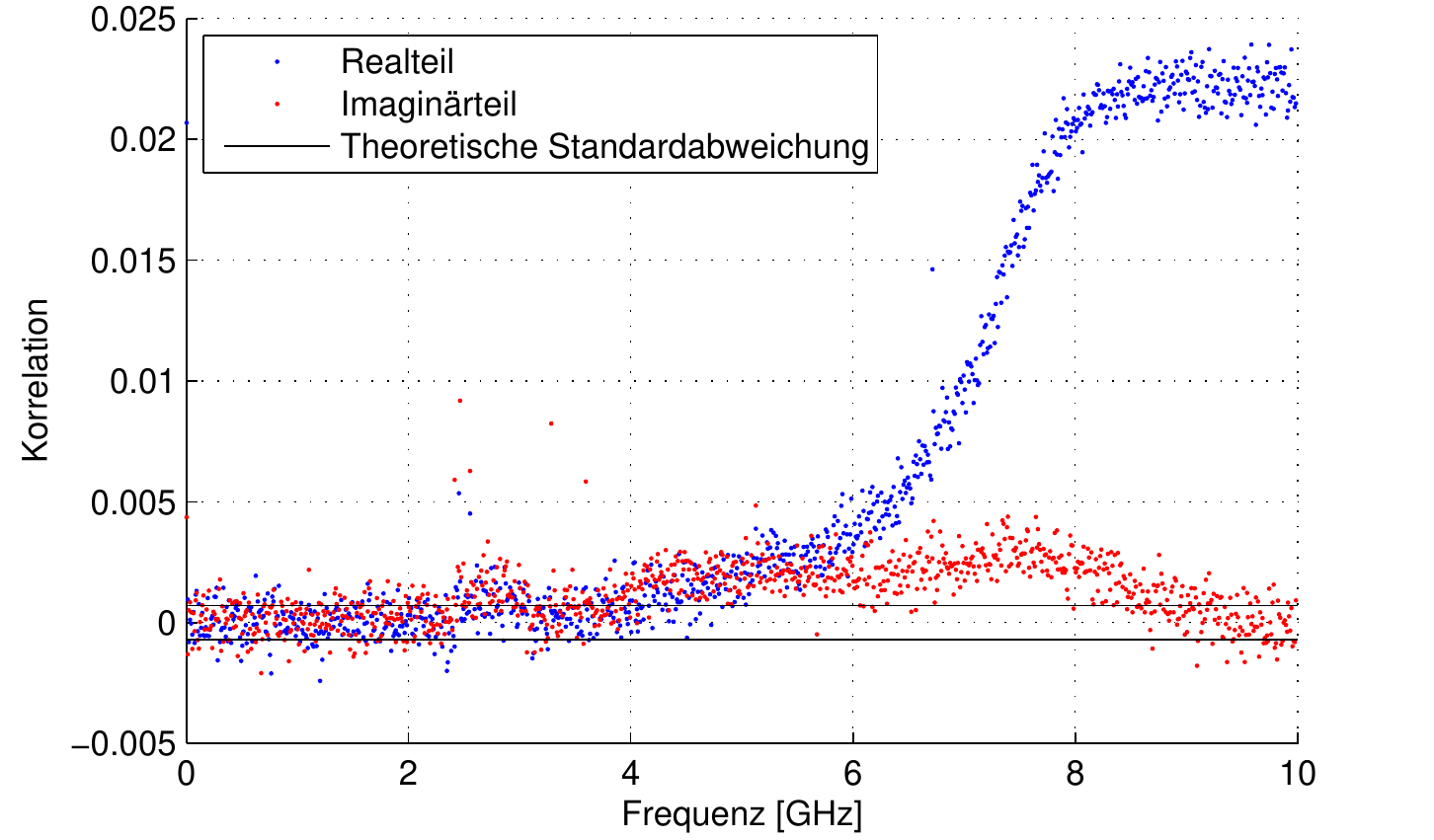}
	\caption{Korrelation benachbarter Frequenzamplituden bei $\Delta f=0{,}1\unit{MHz}$ Auflösung.
Die Werte sind aus $4\ee9$ Datenpunkten berechnet und jeweils am Ende über $10\unit{MHz}$ gemittelt. Im relevanten Bereich bis $1\unit{GHz}$ ist noch kein Trend erkennbar.}
	\label{fig:korrelation}
\end{figure}

Das beobachtete Verhalten war in allen Fällen ähnlich (Abb.~\ref{fig:korrelation}):
Ab Frequenzen über $2\unit{GHz}$, wo der Detektor anfängt stark nichtlinear zu werden, nahmen die Korrelationen deutlich zu.
Bei Frequenzen über $6\unit{GHz}$ wo die Signalamplitude selbst nur noch sehr schwach war und Leck-Effekte kleinerer Frequenzen viel Einfluss nehmen, wurden die Korrelationen signifikant hoch bis hin zu Werten von $0{,}2$ bei $1\unit{MHz}$ Auflösung.

Interessant war nun die Abhängigkeit von der DFT-Auflösung.
Für feinere Frequenz\-auflösungen nahmen alle untersuchten Korrelationen in etwa linear ab, obwohl die einzelnen Frequenzen dabei ja näher zusammenrücken.
Das gibt aber insofern Sinn, dass der Leck-Effekt aus einer dritten Frequenz reduziert wird und die relative Flachheit des Spektrums im Bereich der Frequenzabstände zunimmt.
Besonders schön wird dabei deutlich, dass das Spektrum tatsächlich sehr viele unabhängige Schwingungen enthält.

Mit einer Frequenzauflösung von $0{,}1\unit{MHz}$ wie letztendlich verwendet, waren die Korrelationen unterhalb von $1\unit{GHz}$ so gering dass sie nur schwer unter den statistischen Fluktuationen erkennbar waren.
Abschätzungen durch lineare Extrapolation lieferten Werte von
\begin{equation}
c(\mathrm{Re},\mathrm{Im}) = 1{,}5\ee{-5}
,\qquad
c(t_i, t_{i+1}) = 4\ee{-5}
,\qquad
c(f_i, f_{i+1}) = 5\ee{-5}
\end{equation}
bei $f=1\unit{GHz}$ und $\Delta f=0{,}1\unit{MHz}$.
Derart geringe Korrelationen haben keinen signifikanten Einfluss auf die Entropie.
Allerdings sind die Ergebnisse mit Vorsicht zu genießen, denn eine verschwindende Korrelation ist noch kein hinreichendes Kriterium für statistische Unabhängigkeit.

\FloatBarrier
\subsection{Einfluss von Funkstrahlung}
\label{sec:funk}
Die Frequenzen der Signalfluktuationen entsprechen elektromagnetischen Ultrakurzwellen ($0{,}03\unit{GHz}$ -- $0{,}3\unit{GHz}$) und Dezimeterwellen ($0{,}3$ -- $3\unit{GHz}$), die zur Übertragung von Funksignalen breite Anwendung finden.
Aufgrund der extrem sensitiven Elektronik des Detektors werden solche Wellen eingefangen und zum gemessenen Signal der Lichtintensität hinzuaddiert.

In den betrachteten Messbereich fallen zahlreiche genutzte Sendefrequenzen, z.\,B. Rundfunk ($87{,}5-108\unit{MHz}$), ISM-Bänder\footnote{Industrial, Scientific and Medical Band, für Anwendungen wie Funk-Thermometer oder Alarmanlagen} (433 / $868\unit{MHz}$), Mobilfunk (0{,}9 / $1{,}8\unit{MHz}$) und WLAN/Bluetooth/Mikrowellenherd ($2{,}4-2{,}5\unit{GHz}$).
Diese Funkwellen sind allgegenwärtig aber unterschiedlich stark.

\begin{figure}[bth]
	\centering
		\includegraphics[width=\textwidth]{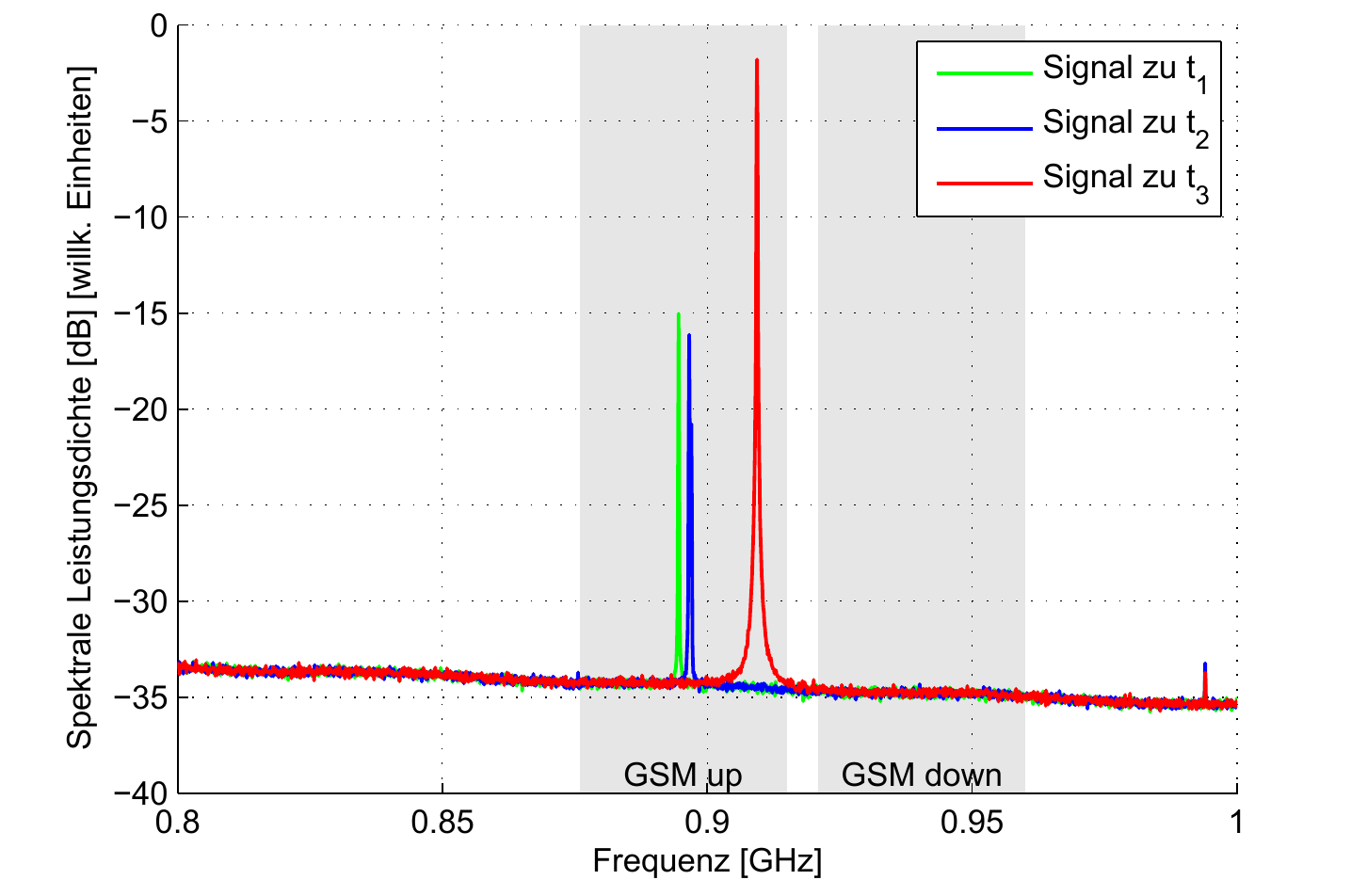}
	\caption{Mobilfunk Spitzen im Detektorsignal zu unterschiedlichen Zeiten.}
	\label{fig:mobilfunk_signal}
\end{figure}

Bei den Messungen stellte sich heraus, dass ausschließlich Frequenzen des GSM\footnote{Global System for Mobile Communications} Mobilfunknetzes detektiert werden konnten, deren Sender sich nahe am Experiment befanden.
Dabei waren aber Signalstärken messbar, die bei einer $0{,}1\unit{MHz}$-Auflösung rund das $10^4$-fache unseres Quantenrauschens erreichen konnten (Abb. \ref{fig:mobilfunk_signal}).

Solange sich Personen mit Handy im Labor befanden war zu einem großen Teil der aufgzeichneten Zeitpunkte mindestens ein Signal erkennbar.
Alle Peakfrequenzen lagen dabei im Uplink-Band zwischen 876 und $915\unit{MHz}$.
Deshalb handelte es sich vermutlich um Signale mit welchen sich die Mobiltelefone bei der Basisstation melden.
Die drei Spitzen in Abb. \ref{fig:mobilfunk_signal} entstammen alle dem D1-Netz.
Der Unterschied der Signalstärken von $14\unit{dB}$ entspricht einem Abstandsverhältnis von 5, was darauf hindeutet, dass die schwächeren Spitzen aus dem kaum abgeschirmten Nachbarlabor stammen könnten.
Die Verbreiterung um rund $5\unit{MHz}$ durch den Leck-Effekt ist bei der höchsten Spitze deutlich zu erkennen.
Wo eine derartige Spitze auftritt, ist das Signal natürlich nicht mehr geeignet um Zufallszahlen zu erzeugen.
Eine Abschirmung des Detektors wäre denkbar, ist aber schwierig wirkungsvoll zu realisieren.
Deshalb werden Sicherheitsmechanismen nötig sein, die beispielsweise unnatürliche Spitzen adaptiv erkennen.
Um das System aber ohne  große Verkomplizierungen sicher zu machen wurde im Experiment der GSM Uplink Bereich plus Sicherheitsbänder von $5\unit{MHz}$ von vornherein ausgeschlossen.
Hierdurch verringerte sich aber auch die Datenrate um rund 5\%.

\FloatBarrier
\section{Diskussion und Schlussfolgerung}
Die Messungen haben gezeigt, dass unser Aufbau die theoretisch erwarteten Eigenschaften gut erfüllt und einwandfreie Zufallszahlen mit einer sehr hohen Rate erzeugen kann.
Durch die Linearitätsmessungen (Kap.~\ref{sec:linear}) wurde nachgewiesen, dass es sich beim Signal um echtes Quantenrauschen handelt und dass der Detektor mindestens bis $1\unit{GHz}$ linear reagiert, unter Umständen auch noch etwas darüber.
Die ideale Strahlleistung liegt bei $5\unit{mW}$, wo die Linearität noch sehr gut ist und das Signal-zu-Rauschverhältnis groß genug.

Die ideale Frequenzauflösung war nicht von vornherein abschätzbar.
Hohe Auflösung haben den Nachteil dass sehr viele Daten getrennt behandelt werden müssen und jede einzelne Statistik für Tests entsprechend geringer ausfällt.
Wie sich aber herausstellte, ist eine hohe Auflösung notwendig um den Leck-Effekt und Korrelationen klein zu halten und um das ursprüngliche Ziel der Methode --~statistisch unabhängige Werte zu erhalten~-- zu erreichen.
Einen guten Kompromiss stellte eine Auflösung von $0{,}1\unit{MHz}$ dar, wodurch je 20\,000 Werte gleichzeitig erzeugt und separat verarbeitet werden und danach entsprechend verkettet werden müssen.

Ein erstaunlich hohes Störpotential von Funksendern wurde in Kap.~\ref{sec:funk} festgestellt, wonach Handystrahlung von nahe gelegenen Geräten das Messsignal um Größenordnungen übertreffen kann.
Das ist besonders für Sicherheitsanwendungen ein relevanter Aspekt und muss bei einer industriellen Realisierung besondere Aufmerksamkeit erhalten.
In unserem Aufbau war es ausreichend, ein begrenztes Frequenzband ($871-920\unit{MHz}$) aus den Daten auszuschließen, aber besser wäre es natürlich solche Einflüsse auf den Detektor von vornherein zu beheben.

Zur Weiterverarbeitung der Frequenzamplituden wurden zwei Arten von Binning untersucht, die beide zur angestrebten Gleichverteilung der Daten führen.
Die Berechnung der bedingten Entropie der erzeugten Verteilungen zeigten, dass der Einfluss des Elektronikrauschens trotz signifikanter Amplitude im Mittel sehr gering ist.
Zum Beweis mathematisch einwandfreier Zufallszahlen musste jedoch der schlimmste Fall betrachtet werden, bei dem aus der Kenntnis des Elektronikrauschens eine größtmögliche Information über das Ergebnis bekannt wird.
Auch für dieses Szenario konnte ein geeignetes Binning gefunden werden.

Die Anzahl der Bins darf für jeden Wert bei mindestens $2^{16}$ liegen, wobei hier noch Raum zu weiterer Optimierung besteht.
Für ein 16 Bit Binning konnte allerdings gezeigt werden dass die bedingte Entropie sehr nahe an der erzeugten Datenmenge liegt.
Das bedeutet, dass die Zahlen in dieser Rohform bereits sehr gute Zufallszahlen sind, und dass die Rate eines anschließenden Hashings fast nur durch den Sicherheitsparameter $\varepsilon$ bestimmt wird.
Bei einer vorsichtigen Hashingrate von 16 auf 14 Bit entstehen so aus Real- und Imaginärteilen auch nach Entfernung der Sicherheitsbänder noch Zahlen mit einer Rate von $25\unit{Gbit/s}$.

Somit wurde ein ausgesprochen leistungsfähiger Zufallsgenerator demonstriert, der quantenmechanische Prinzipien nutzt.
Dabei sind sowohl Qualität als auch Erzeugungsrate bei den höchsten erstrebenswerten Standards angelangt, so dass als Herausforderung nur noch die preisgünstige und benutzerfreundliche Umsetzung der vorgestellten Prinzipien verbleibt.

\newpage
\bibliographystyle{unsrtdin}
\bibliography{ba}

\newpage
\section{Danksagung}
Allem voran möchte ich Prof. Dr. Gerd Leuchs und Dr. Christoph Marquardt danken, dass sie dieses Projekt durch ihre herausragende Organisation der Arbeitsumgebung am MPL möglich gemacht haben.
Sie haben das beste Umfeld geschaffen, das man sich für ein solches Projekt wünschen konnte.

Für die direkte Betreuung meiner Aufgaben möchte ich mich bei Christian Gabriel bedanken.
Seine Unterstützung von Anfang bis Ende hat maßgeblich dazu beigetragen, der Arbeit eine Richtung zu geben und das Projekt erfolgreich durchzustehen.
Seine Erfahrung und Hilfestellungen beim Experimentieren waren unerlässlich.

Besonderer Dank gilt auch Dr. Christoffer Wittmann, der mir stets bei theoretischen Fragen zur Seite stand.
Seine Schlüsselideen haben das Projekt maßgeblich geprägt.
Christoffer hat mir sehr geholfen, Klarheit über entscheidende Konzepte zu gewinnen und Hürden zu überwinden.

Weiterhin möchte ich mich bei Dr. Metin Sabuncu für seine Unterstützung bei den Messungen bedanken und bei Prof. Dr. Elanor Huntington für ihre Mitarbeit am Projekt und die Bereitstellung ihres einzigartigen Detektors.
Den Gruppenmitgliedern der QIV möchte ich für die äußerst angenehme und auflockernde Arbeitsatmosphäre danken.
Die Zeit in der Gruppe war stets kurzweilig und inspirierend.
Rat und Hilfestellung waren immer schnell zu finden.

Schließlich danke ich allen, die sich für Anregungen und Korrekturen dieser Arbeit Zeit genommen haben, insbesondere Stefan Berg-Johansen, Imran Khan, Quirin Spreiter und natürlich meinen Betreuern.

\newpage
\noindent
Hiermit erkläre ich, dass ich die Arbeit selbstständig angefertigt und keine anderen als die angegebenen Hilfsmittel verwendet habe.\\
\vspace{\baselineskip}
\begin{flushright}
Erlangen, 22. September 2011\\
\vspace{5\baselineskip}
Bastian Hacker
\end{flushright}

\end{document}